\documentclass[smallextended]{svjour3}       
\smartqed  

\usepackage{graphicx}
\usepackage{dirtytalk} 
\usepackage{soul} 
\sethlcolor{pink}

\usepackage[title]{appendix} 

\usepackage[lflt]{floatflt} 

\begin{document}

\title{Comparing the Results of Replications in Software Engineering
}


\author{Adrian~Santos  \and Sira~Vegas \and Markku~Oivo \and Natalia~Juristo}


\institute{A. Santos and M. Oivo \at  University of Oulu, Finland \\ \email{\{adrian.santos.parrilla,markku.oivo\}@oulu.fi}           
           \and
           S. Vegas and N. Juristo \at Universidad Polit\'ecnica de Madrid, Spain \\ \email{\{svegas,natalia\}@fi.upm.es}}

\date{Received: -/- / Accepted: -/-}
\maketitle

\begin{abstract}

\textit{Context:} It has been argued that software engineering replications are useful for verifying the results of previous experiments. However, it has not yet been agreed how to check whether the results hold across replications. Besides, some authors suggest that replications that do not verify the results of previous experiments can be used to identify contextual variables  causing the discrepancies. \textit{Objective:} Study how to assess the (dis)similarity of the results of SE replications when they are compared to verify the results of previous experiments and understand how to identify whether contextual variables are influencing results. \textit{Method:} We run simulations to learn how different ways of comparing replication results behave when verifying the results of previous experiments. We illustrate how to deal with context-induced changes. To do this, we analyze three groups of replications from our own research on test-driven development and  testing techniques. \textit{Results:} The direct comparison of p-values and effect sizes does not appear to be suitable for verifying the results of previous experiments and examining the  variables possibly affecting the results in software engineering. Analytical methods such as meta-analysis should be used to assess the similarity of software engineering replication results and identify discrepancies in results. \textit{Conclusion:} The results achieved in baseline experiments should no longer be regarded as a result that needs to be reproduced, but as a small piece of evidence within a larger picture that only emerges after assembling many small pieces to complete the puzzle.

\end{abstract}

\section{Introduction}
\label{sec:introduction}

Replication is a key mechanism in experimentation~\cite{gomez2010replications}. If an experiment is not replicated, there is no way to distinguish between results that happen by chance (the observed event occurred accidentally) and results that, in reality, do exist~\cite{menzies2016perspectives}. Experiments need to be replicated several times before they can produce an established piece of knowledge~\cite{shadish2002quasi}. Each new replication that is run strengthens previous evidence~\cite{gomez2014understanding}.

However, only a small number of replications are performed in software engineering (SE). In a mapping study about replication in SE, da Silva et al.~\cite{da2014replication} found that each experiment is replicated 1.8 times on average. Complications in performing and publishing replications are the main reasons cited for this low average. The major issues identified by Magalhanes et al.~\cite{de2015investigations} regarding why SE replications are difficult to run are a lack of understanding about what a replication is, what a successful replication is, what types of replications there are and their functions, and how replications should be performed.

This research focuses on the problem of what constitutes a successful replication. This is essential for both researchers and developers, as the only way to generate sound evidence about SE---that can be reliably used by developers---is that researchers run successful replications. To tackle this problem, we need to first understand the goal of replications in SE.

It has been argued that the purpose of replications in SE is to verify the results of previous experiments \cite{de2015investigations}\cite{da2014replication}\cite{bezerra2015replication}. Phrases such as \textit{...replication is the repetition of an experiment to double-check its results...} \cite{juristo2009using}, {\textit{...a replication is a study that is run,..., whose goal is to either verify or broaden the applicability of the results of the initial study...} \cite{shull2004knowledge}, \textit{...[replication is] to repeat the study,..., to see if the original result is stable...} \cite{miller2005replicating}, or \textit{...the aim of replication is to verify results...} \cite{gomez2014understanding} abound in the literature \cite{de2015investigations}. 

It has not yet been agreed how to check whether or not  the results hold across replications \cite{de2015investigations}\cite{shepperd2018role}. Some authors in SE (e.g., \cite{brooks2003replication}) suggest that replications verify the results of baseline experiments if both achieve a similar statistical significance of results. Others (e.g., \cite{jorgensen2016incorrect}) suggest that researchers should focus on the replication of effect size rather than statistical significance, since effect size and significance are not necessarily related\footnote{Refer to \cite{camerer2018evaluating} for an overview of the different approaches that have been proposed in different branches of science to check whether or not the results hold across the replications.} \cite{ellis2010essential}. In spite of their intuitiveness, such approaches overlook the fact that both $p$-values and estimated effect sizes are affected by the sample used in the baseline experiment and the replications (a phenomenon known as \textit{sampling error}). Thus, the use of $p$-values or effect sizes to evaluate the similarity of results across a series of replications may lead to misleading conclusions, with the sample used possibly masking  the real results of the replications (i.e., for the population).

Mature experimental disciplines, such as medicine, recommend the use of meta-analysis~\cite{higgins2011cochrane}\cite{borenstein2011introduction}\cite{whitehead2002meta}\cite{cumming2014new}\cite{biondi2016umbrella}. Some authors have also proposed its use in SE \cite{shepperd2018role}. To the best of our knowledge, however, there has been no previous assessment in SE of how these approaches perform. In view of this, we aim to answer the following \textbf{research question}:

\begin{description}

    \item[\textbf{RQ1}] Which approach should be used to assess whether SE replications verify the results of previous experiments?

\end{description}

	This research question has been further divided into:

\begin{itemize}

    \item \textbf{RQ1.1} Are p-values and estimated effect sizes suitable for verifying the results of previous experiments?

    \item \textbf{RQ1.2} Should meta-analysis be used to verify the results of previous experiments?

\end{itemize}

Additionally, some authors argue that replications that do not reproduce the results of previous experiments may help to elicit the contextual variables underlying the differences of results \cite{juristo2009using}. Contextual variables (e.g., the task being developed, the length of the experimental session, the type of subject being used, etc.) may influence the effect size being estimated in each replication~\cite{basili1999building} (i.e., also referred to the effect size in the population~\cite{cumming2013understanding} or the true effect size~\cite{menzies2012special}). 

The SE context is very complex, which means that there are multiple variables that may be impacting the results of experiments \cite{basili1999building},\cite{wohlin2012experimentation}. Some of these variables may have gone unacknowledged, provoking that even slight---possibly unintentional---changes across  replications (e.g., changing the programming language from Java to C++) may alter the true effect size  estimated in each replication. This leads us to our second research question:

\begin{description}

    \item[\textbf{RQ2}] How can we discover if (un)intentional contextual modifications have altered the true effect size and properly interpret results?

\end{description}

To answer these research questions, we first investigate what has been written about aggregation methods and context in SE. Then, we apply simulation to learn how $p$-values and effect sizes behave when the replications are estimating an identical true effect size. In view of the results of these simulations, we claim that $p$-values and effect sizes are unsuitable for verifying the results of previous experiments. We then apply simulation again to illustrate how meta-analysis behaves. According to the results, we propose that replications need to go hand-by-hand with meta-analysis \cite{borenstein2011introduction}. Later, we discuss how to discover and deal with possible context-induced alterations of effect size with meta-analysis. Finally, we illustrate all these issues with three different groups of replications from our own research on test-driven development and testing techniques.

The main \textbf{contributions} of this article are that it illustrates: why $p$-values and effect sizes are unsuitable for verifying the results of replications of SE previous experiments, why meta-analysis should be used instead, and how to detect and deal with context-induced alterations of true effect size so that results can be properly interpreted.

In view of this, we provide several \textbf{take-away messages}. On the one hand, p-values and effect sizes should not be used to verify the results of previous experiments in SE, since most of them have small sample sizes and very often estimate small true effect sizes. On the other hand, the results of a baseline experiment should not be regarded as something to be reproduced, but as a small piece of evidence to be assembled with other pieces to provide a mature piece of knowledge. Therefore, we recommend using meta-analysis to verify the results of SE experiments. More precisely, random-effects meta-analysis, since heterogeneity might materialize. In order to know if contextual modifications have altered the true effect size, the possible influence of sample variables should always be checked. However, configuration variables should only be explored when heterogeneity does materialize. Finally, limitations of the statistical techniques used might be acknowledged: in case the value for heterogeneity is inaccurate, more replications are needed; sample heterogeneity can only be checked when raw data is available; and causality cannot be derived from the sample and configuration variables explored, even if a statistical relationship has been identified.

These results are useful for empirical software engineers and software developers. They suggest that empirical software engineers should be aware of the limitations of using p-values and effect sizes for checking the results of replications, acknowledge the importance of using meta-analysis, and understand the importance of context and its implications when running replications. On the other hand, software developers should take into consideration the method used to check the results of previous experiments when using the evidence generated by empirical software engineers, and learn to understand and take into consideration the impact of context in the evidence generated by empirical software engineers.
	
This article does not intend to discourage SE experimenters from conducting replications. Rather our aim is to convey the idea that replications should not focus on the reproducibility of single results, but that experiments---and replications---should be regarded as just small pieces within a larger picture that only makes sense when they have all been put together.

This paper has been organized as follows. In Section~\ref{background}, we report the background of this study. In Section~\ref{related_work}, we present related work. In Section~\ref{research_method}, we outline the research method of our study. In Section \ref{checking}, we answer RQ1.1 by conducting a series of simulations to show why p-values and effect sizes may not serve to verify the results of previous experiments, and summarize the results of our simulations. Then, in Section~\ref{proposal}, we answer RQ1.2 by conducting a new simulation to illustrate how meta-analysis behaves, and propose its use in SE. In Section \ref{heterogeneity}, we answer RQ2 by showing how to use meta-analysis to detect and deal with replications estimating a different effect size than the experiment. Afterwards, in Section \ref{examples}, we show three real-life groups of replications of our own research where we conduct different types of replications, and provide a discussion about the implications of our study. We discuss the threats to the validity of our study in Section \ref{threats}. Finally, we conclude and present future work in Section \ref{conclusion}. The simulations and analyses run in this research can be found at https://github.com/GRISE-UPM/Comparing-Results-Replications.

\section{Background}
\label{background}

Here, we discuss the main issues that affect SE experiments: sampling error (Section~\ref{background_1}) and context (Section~\ref{background_2}).

\subsection{Sampling Error}
\label{background_1}

Suppose that a development method (test-driven development---TDD \cite{beck2003test}) slightly outperforms another development method (its reverse approach, iterative test-last---ITL) in terms of external quality (measured as the percentage of test cases that successfully passed from a battery of test cases that we developed to test the solutions of the participants) in a specified population of developers (the population of all Finnish developers) using the Java programming language\footnote{We use this example for illustrative purposes, although we are unlikely to ever be able establish this fact, unless we conduct a large enough number of experiments so as to sample the whole population \cite{ioannidis2005most}\cite{cumming2013understanding}\cite{camerer2018evaluating}.}. Suppose that, within this population of Finnish developers, the average quality with TDD is equal to 58, while the quality achieved with ITL is equal to 50. Assuming that, in both  cases, quality follows a normal distribution $\mathcal{N}(50,10^2)$ for ITL and $\mathcal{N}(58,10^2)$ for TDD, Figure \ref{figure_population} shows the two distributions. Their means are represented by two vertical dashed lines.

\begin{figure}[h!]  
  \caption{Population: ITL vs. TDD.}
  \label{figure_population}
  \centering
    \includegraphics[width=9cm,keepaspectratio]{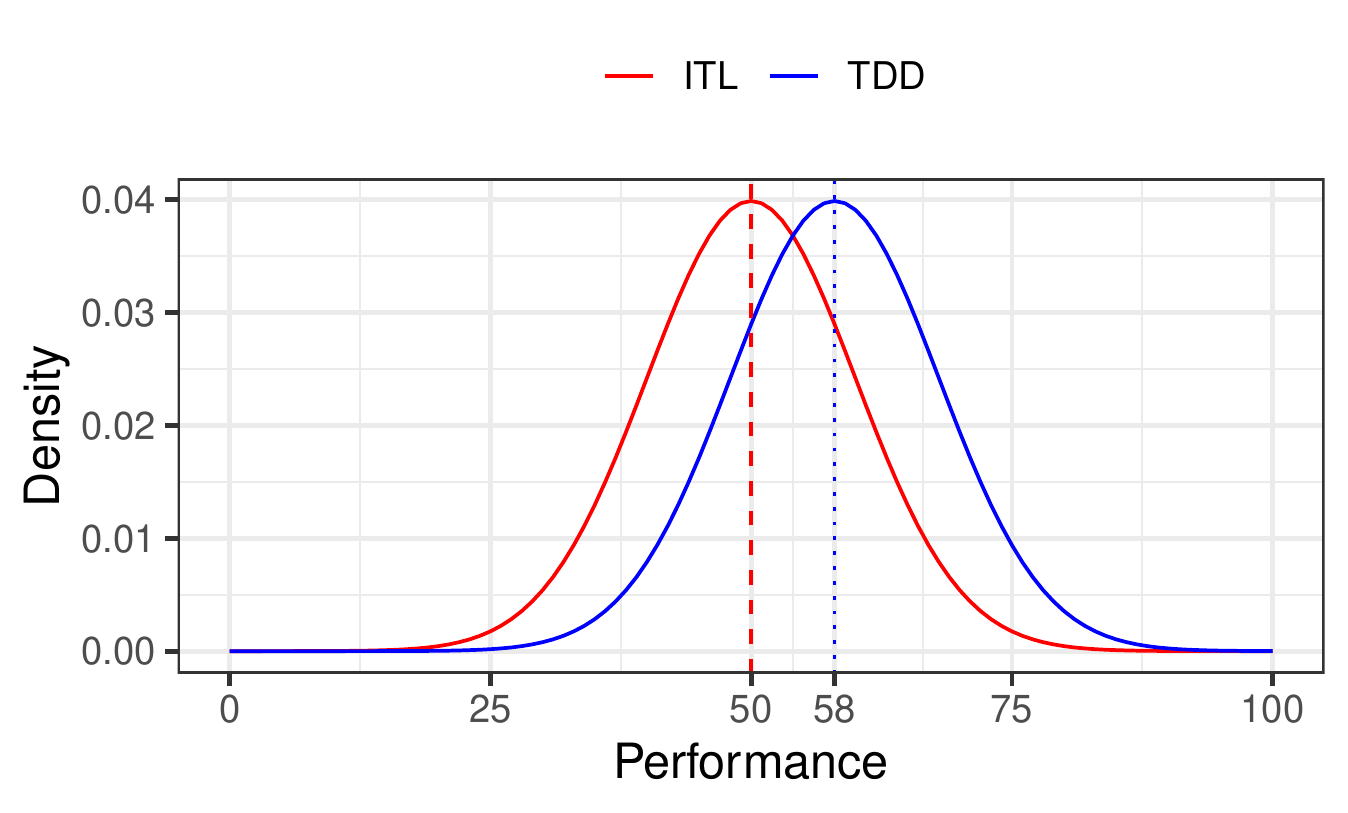}
\end{figure}

As Figure \ref{figure_population} shows, even though the average quality achieved with ITL is equal to 50, some developers achieve lower, and some higher, quality scores. But when averaging out their quality scores, the population mean is equal to 50. 

Suppose that we run a baseline experiment---Experiment 1--- by means of simulation. For this purpose, we sample 20 data-points from each normal distribution to simulate an experiment of 40 developers. Figure \ref{figure_productivity} shows the violin-plot and box-plot for the performance scores reported in the results of the simulated experiment. Table \ref{descriptive_simulation} shows the respective descriptive statistics (sample sizes, means, standard deviations, and medians).

\begin{figure}[h!]  
  \caption{Violin-plot and box-plot: ITL vs. TDD.}
  \label{figure_productivity}
  \centering
    \includegraphics[width=9cm,keepaspectratio]{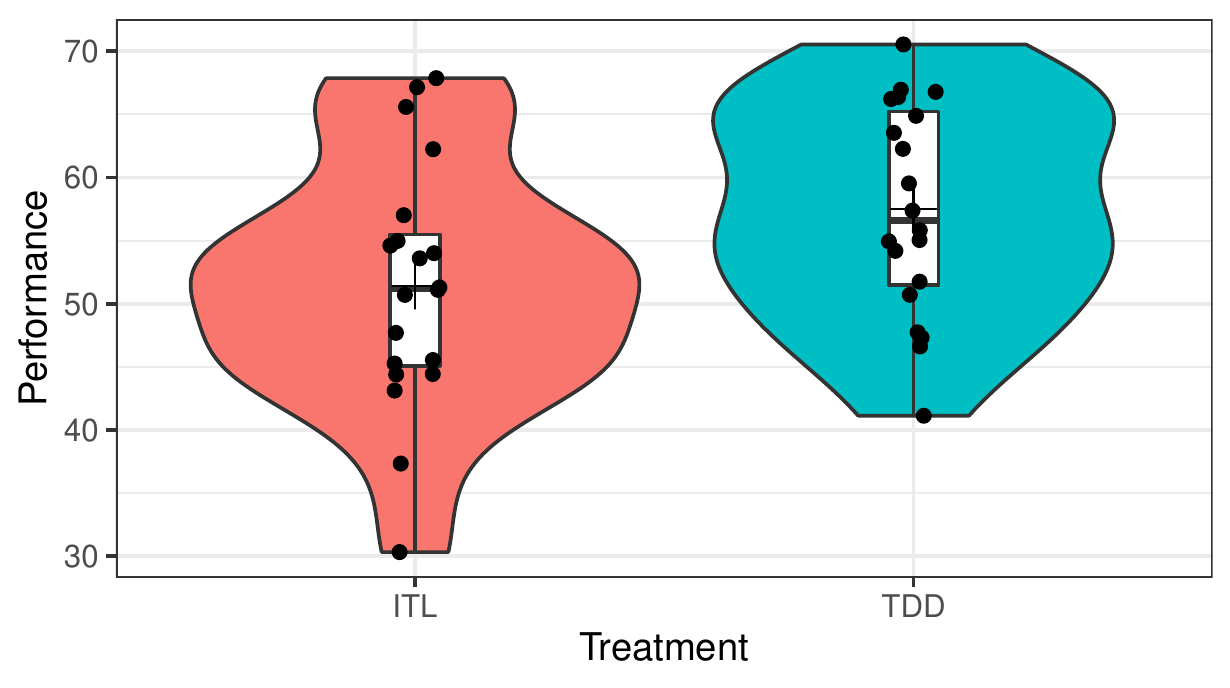}
\end{figure}

\begin{table}[h!] \centering 
  \caption{Descriptive statistics: ITL vs. TDD} 
  \label{descriptive_simulation} 
\begin{tabular}{ccccc} \hline \hline 
\textbf{Method} & \textbf{N} & \textbf{Mean} & \textbf{Std. Dev} & \textbf{Median} \\ \hline 
ITL & $20$ & $51.42$ & $9.73$ & $51.20$ \\ 
TDD & $20$ & $57.49$ & $8.30$ & $56.60$ \\ \hline 
\end{tabular} 
\end{table} 

As Table \ref{descriptive_simulation} shows, the mean performance achieved with ITL ($M=51.42$) is lower than the mean performance achieved with TDD ($M=57.49$). Contrariwise, the standard deviation with ITL ($sd=9.73$) is larger than the standard deviation with TDD ($sd=8.30$). 

The t-test shows that we can reject the experiment null hypothesis of equality of means (t(38)=2.123, p=0.0403). Additionally, a Cohen's d of 0.67\footnote{Cohen's~d$=\frac{57.49-51.42}{\sqrt{(9.73^2+8.30^2)/2}}=0.67$.}---medium \cite{borenstein2011introduction}---materialized in the experiment. Note that its 95\% CI spans (0.03, 1.31), which suggests that,  given the data of the experiment, a wide range of other true effect sizes---from small to large \cite{borenstein2011introduction}---other than 0.67 are also possible in the population. Finally, as the 95\% CI does not cross 0, then we can claim that the effect size is statistically significant.

Notice that the standard deviations, means, and effect size of the simulated experiment were not identical to the respective statistics for the population (i.e., a $\sigma(ITL)=\sigma(TDD)=10$, $\mu(ITL)=50$, $\mu(TDD)=58$, and Cohen's d=0.8, respectively)\footnote{Note that the results of experiments with larger sample sizes are more like the findings for the population.}. This is known as sampling error. It occurs when the statistical characteristics of a population are estimated from a sample (subset) of that population. Since the sample does not include all members of the population, the statistics of the sample and the population tend to differ. And, therefore, different samples vary because they contain different members of the population.

If we simulate N experiments (i.e., replications) from these distributions, they will all be estimating a true Cohen's d of 0.8. But the effect sizes and $p$-values of these experiments will fluctuate \cite{cumming2013understanding} around the true effect size. In other words, p-values and effect sizes are expected to be variable due to sampling error, even though they are all estimating the same true effect size (i.e., an effect size of 0.8 and a statistically significant $p$-value---as the true effect size is not 0)\footnote{When we refer to \say{replications that estimate an identical true effect size}, we do it in the same terms as Borenstein et al.~\cite{borenstein2011introduction}, where the authors refer to a situation in which all factors that could influence the effect size are the same in all the studies, and thus, the true effect size is the same, since all differences in the estimated effects are due to sampling error.}.

At the same time, the smaller the true effect size being estimated, the larger the variability. Finally, the sample size of the baseline experiment and the replications also affect this variability: the smaller the sample sizes are, the larger the expected variability of $p$-values and effect sizes will be, and vice versa\footnote{Note that the experimental design does not affect the value of the estimated effect size, it influences the ability to detect the true effect size, which is known as power.}. \cite{cumming2013understanding}\cite{button2013power} \cite{ioannidis2008research}. As SE experiments are typically small (55 observations on average, the median being 34~\cite{dybaa2006systematic}), their results are largely dependent upon the results of just a few participants (i.e., a small sample participating in the experiment). On the contrary, if SE experiments were larger, the oscillation around the true effect size (i.e., of the population) would be smaller. This is because the larger the sample size is, the less dependent the results of the experiment are on the outcomes of just a few participants \cite{cumming2013understanding}.

\subsection{The SE Context}
\label{background_2}

Note that the development approaches of the experiment described in the previous section cannot be applied in a vacuum. At least one experimental task needs to be used so the participants exercise the development approaches. Suppose that Experiment 1 was performed with an easy task, and we then ran a replication---Experiment 2---with an extremely difficult task. Intuitively, tasks with different complexity may lead to different true effect sizes: while developers set an easy task may achieve different quality scores with each development approach (depending upon the strengths and weaknesses of each development approach), the developers set an extremely difficult task may fail to achieve any quality at all with any development approach---as none of developers may manage to write a single line of code with either ITL or TDD. If we repeat  Experiment 1 and Experiment 2 several times, we may learn that the true effect size in the population with the easy task was positive (TDD outperformed ITL), but that the true effect size in the population with the extremely difficult task was equal to 0 (as not a single participant managed to write a single line of code with either ITL or TDD). Thus, intuitively, if we run two different replications, each with a different experimental configuration (i.e.,  if we make purposeful changes in the experimental tasks or session lengths across replications, for example), each replication may be---at least potentially---estimating a different true effect size \cite{whitehead2002meta}\cite{cumming2014new}\cite{borenstein2011introduction}.  

The same may also hold  if we evaluate different types of participants  across the replications. For example, if the participants in the baseline experiment have 10 years of programming experience and the participants in the replication have no experience at all, the true effect size to be estimated in the baseline experiment could be positive (again, supposing that TDD outperformed ITL), and the true effect size  to be estimated in the replication could be 0---as none of the developers may be able to write a single line of code.

Note that while changes made across the replications may alter the true effect size being estimated in each replication, this will only be the case if the changed variable(s) has(have) an impact on the true effect size. Likewise, the fact that replications are identical does not imply that all the replications estimate an identical effect size. In particular, as many characteristics of the replications or the participants may impact the true effect size being estimated in SE  \cite{wohlin2012experimentation}\cite{basili1999building}, unacknowledged characteristics of the replications may also influence the true effect sizes being estimated in each replication \cite{higgins2011cochrane}\cite{whitehead2002meta}.

The SE context is very complex~\cite{juristo2011role}, and we still do not know which variables are impacting the results. Therefore, any changes made may easily go unnoticed.

Accordingly, the problem related to sampling error and the influence of context in experimentation generally is magnified in SE due to typical small sample sizes and the complexity of the context of the experiments.

\section{Related Work}
\label{related_work}

According to SE researchers, replication has to do with verifying the results of previous experiments~\cite{kitchenham2008role}\cite{gomez2014understanding}. Therefore, this section commences with a  bird's-eye view of the approaches used in SE to verify the results of previous experiments. Then, we discuss the role of simulations in experimental SE, this being the approach followed in this research. Finally, we highlight the importance of the SE context, which is another concern of this paper.

As discussed by de Magalhaes et al. \cite{de2015investigations} and Shepperd et al.~\cite{shepperd2018role}, one key issue related to replication in SE addresses the different procedures that can be followed to increase the chance of running a successful replication. Shepperd et al.~\cite{shepperd2018role} highlight that this issue has been addressed by only a few studies and superficially. In a systematic mapping study that we conducted \cite{adrisms}, we identified the techniques used to aggregate the results of groups of replications in SE: $p$-values, and more recently, effect sizes and their 95\% confidence intervals (95\% CIs), prediction intervals (95\% PIs) and meta-analysis \cite{wohlin2012experimentation}\cite{juristo2013basics}.


A \textit{$p$-value} is reproduced when the replication and the baseline experiment output a similar value (i.e., if the $p$-values of the baseline and the replication are statistically significant\footnote{Note, however, that the true effect size is just as important, and, in actual fact, given the problems with questionable research practices and publication bias, a small experiment will most likely overestimate the effect size because it is impossible for a small effect size to be significant.}). An \textit{effect size} is reproduced when the replication and baseline experiment output a similar value. The \textit{effect size 95\% CI} of a baseline experiment is reproduced if the 95\% CI of the estimated effect sizes of the replication and the baseline experiment overlap. \textit{Prediction intervals} (i.e., 95\% PIs) are one of the latest methods proposed to assess whether the results hold across replications \cite{patil2016should}. They are a statistical technique for predicting the range of effect sizes that researchers should expect in a replication study. This way, it is possible to form a global view of whether the results of the two are consistent. Note that PIs operate in the same manner as CIs, albeit with respect to an individual future observation given what has already been observed. \textit{Meta-analysis}~\cite{borenstein2011introduction} pools together the results of different replications into a joint result. It is commonly recommended in mature experimental disciplines, such as medicine \cite{higgins2011cochrane}, with the aim of building on the results of previous experiments.  


Although different approaches are used in the experimental SE literature to assess whether the results hold across replications, there is, to the best of our knowledge, no previous work that formally compares their performance.

Simulations are not new to experimental SE. They have previously been used to simulate the behaviour of testing techniques~\cite{duran1984evaluation} \cite{ntafos1998random}, to illustrate the effect of researcher and publication bias~\cite{jorgensen2016incorrect}, or to provide guidelines on how to report them~\cite{travassos2016experimentation}. Note that other researchers, like Patil et al.~\cite{patil2016should}, have run simulations in the field of Psychology, to investigate the performance of $p$-values and effect sizes, but there are no works specific to SE (i.e., where the parameters of the simulations are set to the specific characteristics of SE).

Finally, the importance of the SE context has been highlighted by several authors. On the one hand, Briand et al. \cite{briand2017case} argue that the applicability and scalability of SE solutions depend largely on contextual factors, and that they vary widely from one context to another. On the other hand, Murphy \cite{murphy2019beyond} stresses the need to build the concept of context in SE: we need to recognize what context is, what form it takes, and we need to be able to better define and implement context. This suggests that context is a key issue when running SE experiments. Its importance is also acknowledged in experimental SE with works that try to establish what the influential context variables are \cite{petersen2009context} \cite{host2005experimental} \cite{badampudi2019contextualizing} or guidelines for reporting empirical studies \cite{runeson2009guidelines} \cite{jedlitschka2008reporting}.

\section{Research Method}
\label{research_method}

We put the approaches that SE researchers use to verify the results of previous experiments (checking $p$-values, effect size magnitudes and signs, or meta-analysis) to the test by means of simulation. The simulation approach that we adopted was modeled on Cumming \cite{cumming2013understanding}. 

We decided to resort to simulations to gain insight into how such approaches may work for verifying the results of previous experiments for two reasons. The first reason is that unless we conduct a large enough number of experiments so as to ultimately sample the whole population, we may never get to know the real shape of the population distributions  \cite{ioannidis2005most}\cite{cumming2013understanding}\cite{camerer2018evaluating}. Simulation, however, can provide this information, giving us access to the true effect size in the population. We can then simulate experiments from this population (i.e., by drawing different samples) and then observe how experiments estimating the same true effect size behave in terms of $p$-values, effect sizes and meta-analysis. The second reason is that through simulation, we can produce visual and statistical evidence (e.g., density plots, box plots and descriptive statistics \cite{cumming2013understanding}\cite{field2013discovering}) to aid the understanding of results.

Continuing with the example given in Section~\ref{background}, we simulate the performance of two development approaches (ITL and TDD) within a specified population of developers (Finnish developers) with respect to a continuous outcome of interest (external quality) with normal distributions\footnote{We are aware that SE data may not be normal \cite{kitchenham2017robust}\cite{arcuri2011practical}. However, we opted to use normal distributions as they are a convenient way of expressing the true effect size in the population in terms of Cohen's d  \cite{borenstein2011introduction}. We decided to express the effect size using Cohen's d because of its common use in SE \cite{kampenes2007systematic}. We discuss the shortcomings of simulating normally distributed data in the threats to validity section.}.

In particular, we simulate several series of experiments in Section~\ref{checking}. Each series simulates 5,000 experiments estimating the same true effect size with the same sample size. We simulate 5,000 experiments, since this is a typical number in Montecarlo simulations \cite{morris2019using}. Different series simulate experiments with different true effect sizes and sample sizes. The selected true effect sizes are 0.2, 0.5 and 0.8, as they correspond to the threshold for small, medium and large effects, respectively \cite{borenstein2011introduction}. The selected sample sizes follow an arithmetic progression from 4 to 148 with a common difference of 16 in line with the values reported by Dyb\aa\ et al.~\cite{dybaa2006systematic}, where a minimum of 5 and a maximum of 136 were used for experiments analyzed with a t-test (the statistical test that we are simulating).

Each experiment is simulated by sampling a specified number of times from each distribution---determined by the number of participants that we want to simulate with each development method. 

An experiment in which the difference in performance between the two development approaches in the population (i.e., the true effect size) is equal to a Cohen's d of 0.8 will be simulated using two normal distributions: ITL will follow a $\mathcal{N}(50,10^2)$ and TDD a $\mathcal{N}(58,10^2)$\footnote{Cohen's d $=\frac{58-50}{10}=0.8$~\cite{cumming2013understanding}.}. To obtain effect sizes of 0.2 and 0.5, TDD will be simulated with a $\mathcal{N}(52,10^2)$ and a $\mathcal{N}(55,10^2)$, respectively.

Once simulated, the experiments are then analyzed individually to discover the size of the oscillation of their $p$-values and estimated effect sizes around the true effect sizes in the population. Intuitively, if there is a large oscillation, the results of a replication may be uninformative about the results in a baseline experiment. Based on the results of the simulations that we ran, we argue why we think that $p$-values and effect sizes are unsuitable for verifying the results of previous experiments. 

In Section~\ref{proposal}, we report two simulations. The first one compares the estimated effect sizes of 5,000 simulated experiments of sample size 144 with the joint estimated effect sizes obtained if the data of the experiments would have been obtained in groups of 4, 8 or 12 different replications instead of a single big experiments.

The second simulation illustrates how meta-analysis behaves. Here, we simulate several series of groups of experiments (instead of single experiments). Each series simulates 5,000 groups of experiments with the same group size, estimating the same true effect size with the same sample size. Different series simulate groups of experiments with different group sizes, true effect sizes and sample sizes. The selected group sizes range from 1 to 12, which are consistent with the values reported in~\cite{adrisms}, ranging from a minimum group size of 3 to a maximum of 12. The selected true effect sizes are 0.2, 0.5 and 0.8, for the same reason as above. Finally, the selected sample sizes are are 4, 36 and 100. They have been chosen in line with the values reported for a t-test in~\cite{dybaa2006systematic} (a mean of 34, ranging from 5 to 136).

We later study how estimated effect sizes vary depending on the value of the true effect size and the size of the group of replications. In view of the results of the simulations, we propose---as researchers in other disciplines \cite{borenstein2011introduction}\cite{whitehead2002meta}\cite{cumming2014new}\cite{biondi2016umbrella} and particularly SE \cite{shepperd2018role} did before us---to use meta-analysis and conclude why it should be used.


Finally,  Section~\ref{heterogeneity}  shows how meta-analysis can detect and address the fact that different replications are estimating different true effect sizes. This is based on the results of a literature review that we ran in the social sciences and on mature experimental disciplines, such as medicine. In this literature review, we identified articles addressing the issue of replication from a statistical perspective \cite{thompson1994pivotal}\cite{makel2012replications}\cite{button2013power}, books about meta-analysis and related topics such as linear mixed models \cite{borenstein2011introduction}\cite{whitehead2002meta}\cite{brown2014applied}, and references discussing the advent of a \say{reproducibility crisis} across the sciences \cite{pashler2012editors}\cite{baker2016there}\cite{maxwell2015psychology}\cite{camerer2018evaluating}. Based on their results, we surmised that different experimental configurations (i.e., different experimental designs, session lengths, experimental tasks, etc., of the replications) or different participant characteristics (e.g., their different experiences, backgrounds, etc.) across replications may alter the true effect size being estimated in each replication. 

In Section~\ref{examples}, we illustrate the advantages of using meta-analysis by showing and analyzing three groups of replications of our own research on test-driven development (TDD) \cite{beck2003test} and testing techniques \cite{myers2011art}. In these groups of replications, we made either: (1) no changes across the replications; (2) inadvertent population changes across the replications; or (3) purposeful configuration changes across the replications in this case by changing their programming environments.

\section{RQ1.1 Are $P$-Values and Estimated Effect Sizes Suitable for Verifying the Results of Previous Experiments?}
\label{checking}

In this section, we show how $p$-values (Section~\ref{checking_p_values}) and effect sizes (Section~\ref{checking_effect_sizes}) may fail to assess the extent to which the results of replications hold or do not hold. Finally, we discuss the results (Section~\ref{findings}).

\subsection{Simulations for $p$-values}
\label{checking_p_values}

Some authors in SE (e.g., \cite{brooks2003replication}) suggest that the results of a baseline experiment are reproduced if both the results of this experiment and a replication are statistically significant. However, this approach has a major shortcoming, namely, $p$-values are affected by sampling error, which in turn is aggravated by small sample sizes and true effect sizes being estimated \cite{cumming2013understanding}. Thus, the probability of results being statistically significant is lower with smaller experiments or estimated true effect sizes  \cite{cumming2013understanding}. We illustrate this idea in Figure \ref{ratio_significant}, where we show the percentage of statistically significant results achieved across 5,000 simulated replications that estimate the same true effect size at different sample sizes (4...148) and true effect sizes (Cohen's d equal to 0.2, 0.5 and 0.8).
\begin{figure}[t!]  
  \caption{Percentage of statistically significant results across 5,000 replications estimating the same true effect size, for different true effect sizes and sample sizes.}
  \label{ratio_significant}
  \centering
    \includegraphics[width=\textwidth,keepaspectratio]{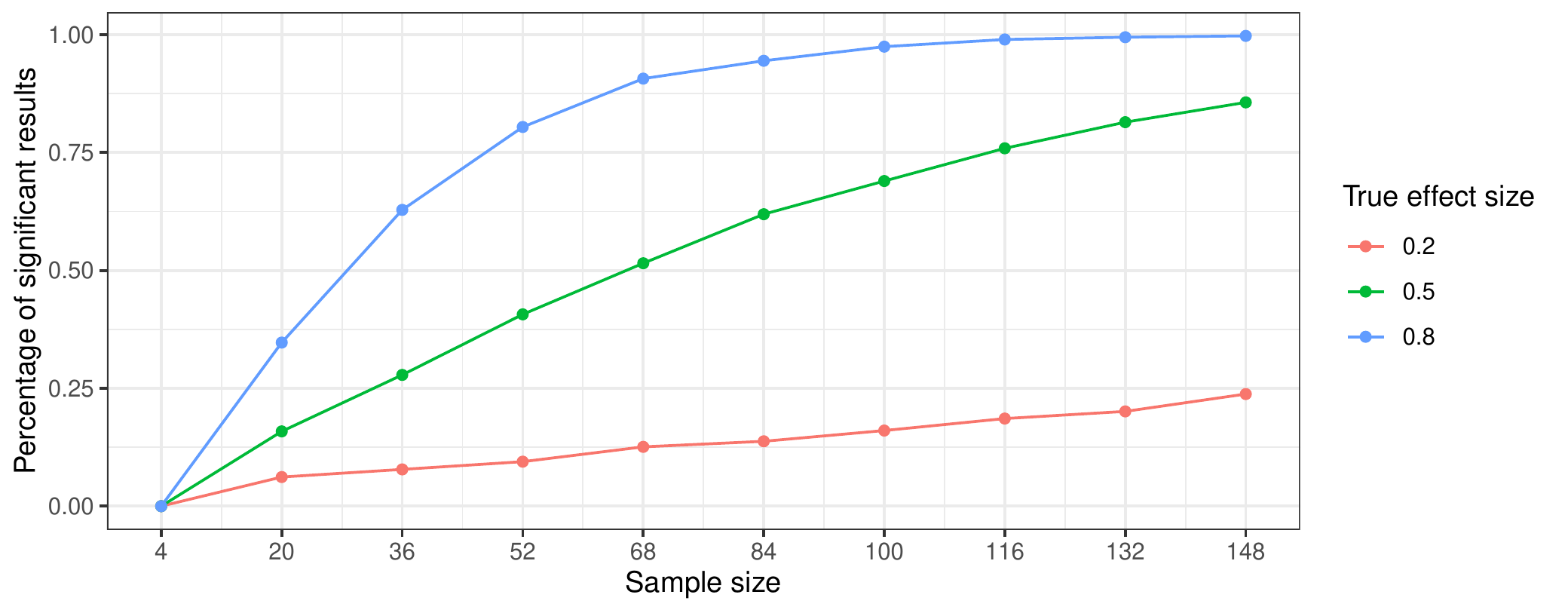}
\end{figure}

As Figure \ref{ratio_significant} shows, the larger the sample size, the greater the probability of the results being statistically significant. This happens for all estimated true effect sizes (0.2, 0.5 and 0.8), as Figure 3 shows increasing lines. Also, the larger the estimated true effect sizes, the greater the probability of the results being statistically significant. This happens for all sample sizes (from 4 to 148). Thus, the probability of getting two, three, four, etc., statistically significant results across identical successive replications is dependent upon the sample sizes of the replications (i.e., the larger the sample sizes, the greater the probability) and the estimated true effect sizes across the replications. 

The relationship between the sample sizes of the replications and their true effect sizes is captured by the concept of statistical power \cite{cumming2013understanding}. The statistical power increases in proportion to the sample size for a given true effect size, where there is a greater probability of detecting that the true effect size is not 0 (i.e., producing a statistically significant result). Thus, the lines of the graph shown in Figure \ref{ratio_significant} can be simply read as the statistical power achieved in AB between-subjects experiments with sample sizes between 4 to 148 and a Cohen's d in the population of 0.2, 0.5 and 0.8.

For example, Figure \ref{ratio_significant} shows that the probability of achieving a statistically significant result in an AB between-subjects experiment with a sample size of 20 and a population  with a Cohen's d of 0.8 (i.e., the top  line) is around 35\% (i.e., the statistical power is equal to 35\%). If, instead of one experiment, we conduct two experiments estimating the same true effect size, each with a statistical power of 35\%, the probability of producing two statistically significant results would be equal to 0.12 ($0.35^2$). If, instead of two, three identical experiments were run, the probability would be equal to 0.04 ($0.35^3$). In general, the probability of achieving N consecutive statistically significant results across N identical replications is equal to \cite{gnedenko2020theory}: 

$$ P(identical)= power^{\#exps} $$

Plugging in the probabilities calculated in Figure \ref{ratio_significant} into the previous formula, and considering groups formed by 2 to 12 identical replications, the probabilities of getting significant $p$-values across all the replications (and, thus, always reproducing the results of previous experiments) can be calculated for all sample sizes and true effect sizes. For example, the probability of reproducing the results of previous experiments in groups of 2 to 12 identical replications for a true Cohen's d of 0.2 at different sample sizes are shown in Figure \ref{identical_results}. Appendix~\ref{reproducing_results} shows the respective figures for a true Cohen's d of 0.5 and 0.8.

\begin{figure}[h!]  
  \caption{Probability that all replications reproduce the results of previous experiments by means of $p$-values in groups of 2 to 12 replications, all estimating a true Cohen's d of 0.2.}
  \label{identical_results}
  \centering
    \includegraphics[width=\textwidth,keepaspectratio]{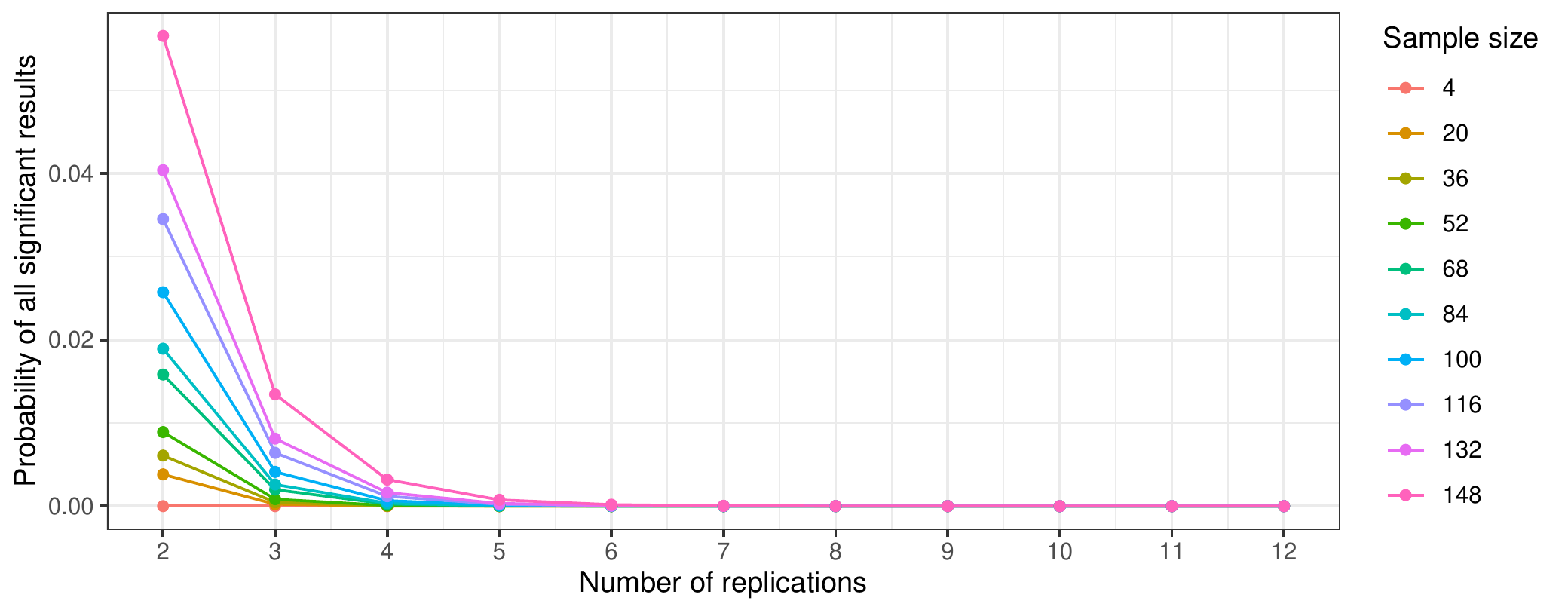}
\end{figure}

As Figure \ref{identical_results} shows, the probability that all replications in a group (containing 2 to 12 replications) reproduce the results of previous experiments in $p$-value terms (i.e., by finding \textit{N} consecutive statistically significant results) is ridiculously small for a true Cohen's d of 0.2---as the probability is less than even 6\% for sample sizes as large as 148 per replication. Although larger probabilities would be achieved at a higher true Cohen's d in the population (as both the probability and the product of the probabilities of finding statistically significant results in each replication would be greater), the probability of reproducing the results of previous replications with the typical sample sizes of SE replications (i.e., around 35 according to Dyb\aa\ et al. \cite{dybaa2006systematic}) is ridiculously small (e.g., a probability of around 0.40 for producing two statistically significant results in two identical replications estimating a true Cohen's d of 0.8 in the population as shown in Figure~\ref{identical_results_large} in Appendix~\ref{reproducing_results}). \\

\noindent\fbox{
  \parbox{\textwidth}{
    \textbf{Key findings}
    \begin{description}
    
    \item[\textbf{KF1}] $p$-values are, due to sampling error, unsuitable for assessing whether or not the results of previous experiments are verified in any SE replication, but especially small ones, which are the most common.

    \item[\textbf{KF2}] $p$-values are especially unsuitable in SE replications estimating small true effect sizes.
        
    \end{description}
  }
}
\\

\subsection{Simulations for Effect Sizes}
\label{checking_effect_sizes}

Some SE researchers (e.g., \cite{jorgensen2016incorrect}) suggest that the results of a baseline experiment are reproduced whenever effect sizes are replicated. Unfortunately, this approach has a major shortcoming, namely, that even exactly identical experiments---especially if they are small \cite{button2013power}\cite{cumming2013understanding}---may throw up conflicting results. Figure \ref{variability_effects} illustrates this point by showing the variability of effect sizes achieved across 5,000 simulated replications, estimating the same true effect size at different sample sizes (i.e., sample sizes between 4 and 148) and true effect sizes (i.e., a true Cohen's d of 0.2, 0.5 and 0.8, from left to right, respectively). We show the true effect size being estimated with a solid black vertical line.

\begin{figure}[h!]  
  \caption{Variability of effect sizes for different sample sizes and true effect sizes (i.e., small, medium, large).}
  \label{variability_effects}
  \centering
    \includegraphics[width=\textwidth,keepaspectratio]{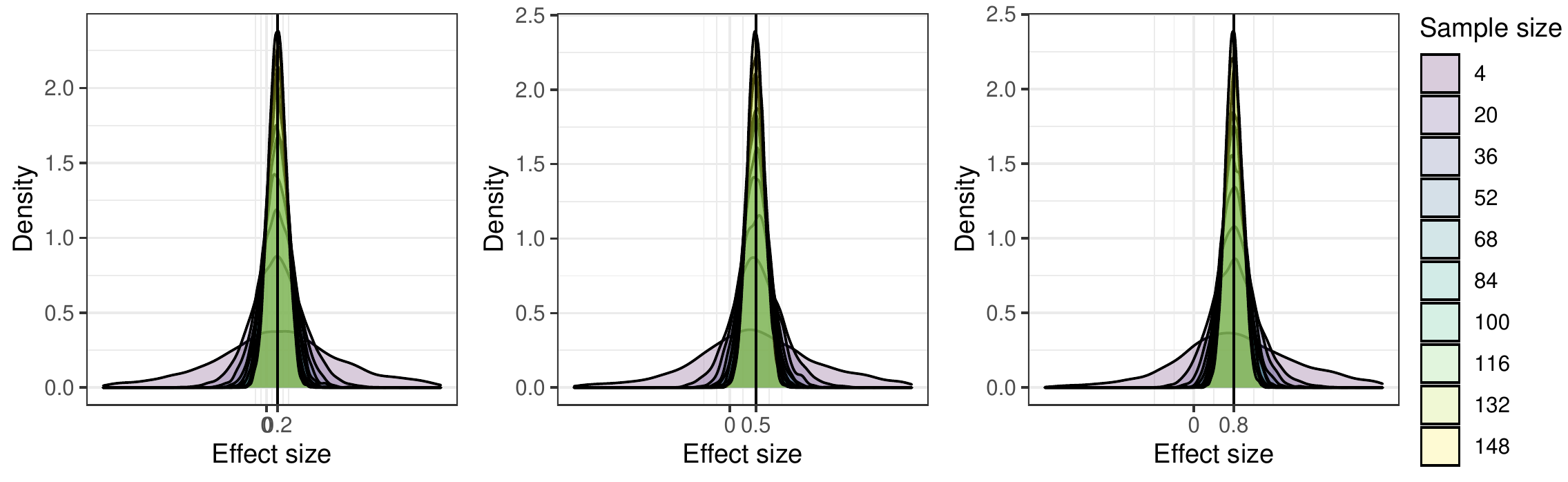}
\end{figure}

As Figure \ref{variability_effects} shows, effect sizes are always variable. Also, independently of the  estimated true effect size in the replications, the variability of the effect sizes increases for smaller sample sizes (notice that the peaks in Figure \ref{variability_effects} are wider and rounder for smaller sample sizes regardless of the estimated true effect size). Thus, a smaller sample size results in a larger variability in effect sizes, which, in turn, leads to a greater probability of it being claimed that the results were not reproduced---when, in actual fact, the results were the same. Table~\ref{confidence_intervals} shows the variation---95\% CIs---in the estimated effect sizes.

\begin{table}[!htbp] \centering 
  \caption{95\% confidence intervals for estimated effect sizes} 
  \label{confidence_intervals} 
\begin{tabular}{@{\extracolsep{5pt}} ccccc} 
\\[-1.8ex]\hline 
\hline \\[-1.8ex] 
True  & Sample & \multicolumn{3}{c}{Estimated effect size} \\ \cline{3-5}
effect size & size & Lower bound 2.5\% & Mean & Upper bound 97.5\% \\ 
\hline \\[-1.8ex] 
$0.2$ & $4$ & $$-$3.55$ & $0.55$ & $5.30$ \\ 
& $20$ & $$-$0.74$ & $0.20$ & $1.16$ \\ 
& $36$ & $$-$0.47$ & $0.20$ & $0.90$ \\ 
& $52$ & $$-$0.34$ & $0.21$ & $0.78$ \\ 
& $68$ & $$-$0.27$ & $0.20$ & $0.69$ \\ 
& $84$ & $$-$0.22$ & $0.20$ & $0.65$ \\ 
& $100$ & $$-$0.21$ & $0.20$ & $0.59$ \\ 
& $116$ & $$-$0.17$ & $0.20$ & $0.57$ \\ 
& $132$ & $$-$0.14$ & $0.20$ & $0.53$ \\ 
& $148$ & $$-$0.13$ & $0.20$ & $0.52$ \\ 
$0.5$ & $4$ & $$-$2.73$ & $0.88$ & $6.24$ \\ 
& $20$ & $$-$0.40$ & $0.52$ & $1.53$ \\ 
& $36$ & $$-$0.14$ & $0.51$ & $1.23$ \\ 
& $52$ & $$-$0.03$ & $0.51$ & $1.07$ \\ 
& $68$ & $0.03$ & $0.51$ & $1.02$ \\ 
& $84$ & $0.07$ & $0.50$ & $0.96$ \\ 
& $100$ & $0.12$ & $0.50$ & $0.91$ \\ 
& $116$ & $0.14$ & $0.50$ & $0.89$ \\ 
& $132$ & $0.17$ & $0.50$ & $0.85$ \\ 
& $148$ & $0.18$ & $0.50$ & $0.85$ \\ 
$0.8$ & $4$ & $$-$1.99$ & $1.36$ & $7.42$ \\ 
& $20$ & $$-$0.08$ & $0.83$ & $1.89$ \\ 
& $36$ & $0.15$ & $0.83$ & $1.60$ \\ 
& $52$ & $0.25$ & $0.82$ & $1.43$ \\ 
& $68$ & $0.32$ & $0.81$ & $1.33$ \\ 
& $84$ & $0.37$ & $0.81$ & $1.30$ \\ 
& $100$ & $0.39$ & $0.81$ & $1.23$ \\ 
& $116$ & $0.45$ & $0.81$ & $1.18$ \\ 
& $132$ & $0.45$ & $0.80$ & $1.17$ \\ 
& $148$ & $0.48$ & $0.80$ & $1.13$ \\ 
\hline \\[-1.8ex] 
\end{tabular} 
\end{table}

Finally, as Figure \ref{variability_effects} shows, the smaller the estimated true effect size is (see the left-hand plot in Figure \ref{variability_effects}, estimating a true Cohen's d of 0.2 in all the replications), the greater the probability of effect sizes being negative will be (as more and more effect sizes are on the left of 0). Thus, with a smaller estimated true effect size, there is a greater probability of there being differences across the signs of the effect sizes in the replications, which, in turn, results in a higher likelihood of results being claimed to be contradictory. \\

\noindent\fbox{
  \parbox{\textwidth}{
    \textbf{Key findings}
    \begin{description}
    
        \item[\textbf{KF3}] SE replications will show variability of effect size magnitudes. Those with small sample sizes---which are very common---will show larger variability.
        
        \item[\textbf{KF4}] The probability of there being conflicting effect size signs is greater for smaller estimated true effect sizes.
        
    \end{description}
  }
}

\subsection{Discussion/Findings}
\label{findings}

In this section, we address RQ1.1: whether $p$-values and/or estimated effect sizes are suitable for assessing the reproducibility of the results of the baseline experiment. Our simulations suggest that they are not. 

Additionally, $p$-values and estimated effect sizes depend upon the sample sizes of the experiments. If the sample sizes are small, this may mask the true effect sizes being estimated. For example, suppose that a  hypothetical population of developers has a true Cohen's d of 0.2. If we run a small baseline experiment with 36 developers estimating the true effect size and observe a Cohen's d of 0.5, this does not imply that the true effect size in the population is equal to a Cohen's d of 0.5. It simply means that, given the large variability of results expected in small sample sizes (according to Table~\ref{confidence_intervals}, the upper bound of the 95\% CI in this situation is 0.90), the best possible estimate of the true effect size given the data of the experiment, is equal to a Cohen's d of 0.5. Likewise, if we observe a  negative Cohen's d (e.g., a Cohen's d of -0.2) for a replication with the same sample size estimating the same true effect size, it does not imply that the baseline experiment and the replication estimate different true effect sizes. These two cases are simply different sides of the same dice (as, according to Table~\ref{confidence_intervals}, the lower bound of the 95\% CI in this situation is -0.47). We should highlight that, although the above two situations are especially likely in small experiments typical of SE \cite{dybaa2006systematic}---where there is expected to be a large variability of results \cite{shepperd2018role}, all SE experiments are subject to this phenomenon. For example, if we run a replication estimating the same true effect size with 148 developers, the 95\% CI still ranges from -0.13 to 0.52 (as shown in Table~\ref{confidence_intervals}).

Exactly the same thing applies for $p$-values. In particular, if we find a statistically significant result in the baseline experiment, and a non statistically significant result  in an identical replication, this does not imply that the replication failed to reproduce the results of the baseline. It simply means that replication did not detect that a true effect size that was not 0 in the population \cite{cumming2013understanding}. Note that, according to Figure~\ref{ratio_significant}, only around 30\% of the experiments run with sample size 36 and estimating a true effect size of 0.5 will output significant results. Although the percentage increases to around 85\% for experiments with 148 developers, it is still a possibility.

Although 95\% CIs are an improvement on $p$-values and estimated effect sizes (the CI for an effect size estimated from a low-powered study will be broad and suggest the results should be interpreted with caution), unfortunately, they do not address all the shortcomings of effect sizes and $p$-values (particularly in the case of small sample sizes). This is because 95\% CIs also depend on the sample size of the experiments (i.e., large and small experiments tend to provide narrow  and wide 95\% CIs, respectively \cite{cumming2013understanding}). For example, we might run a baseline experiment estimating a true effect size of 0.5 with 1000 developers and get a narrow 95\% CI (Cohen's d=0.52, 95\% CI(0.64,0.39)) and then run a replication with 40 developers (estimating the same true effect size) and get an effect size way outside the bounds of the 95\% CI of the baseline experiment (Cohen's d=-0.42, 95\%CI (0.20,-1.05)). The fact that the 95\% of the effect size of the replication does not overlap with the 95\% CI of the baseline experiment does not imply that the replication failed to reproduce the results of the baseline. It simply means that the sample sizes of both experiments also had a bearing on the results. The opposite is also true: if a baseline experiment estimating a true effect size of 0.5 is run with 40 developers, then a wide 95\% CI is likely to materialize \cite{cumming2013understanding} (e.g., Cohen's d=0.68 and  95\% CI =(1.31,0.06)), and the effect sizes of any prospective replications are likely to fall within the bounds of the 95\% CI of the baseline, regardless of their sample sizes or estimated true effect sizes of. For example, a Cohen's d of a replication estimating a true effect size of 0.8 run with 1000 developers may be 0.74 and a 95\%CI (0.87,0.61). This does not mean that the replications successfully reproduced the results of the baseline, as each replication may ultimately be estimating a completely different true effect size. This simply means that, as a byproduct of the small sample size of the baseline experiment, all the effect sizes of the replications may fall within the bounds of its 95\% CI, regardless of the true effect sizes that they are estimating. 

The findings for 95\% CIs also apply to 95\% prediction intervals. In particular, as small experiments tend to be associated with wide 95\% PIs \cite{patil2016should}, then the results of prospective replications are then likely to fall within the bounds of the 95\% PI of a small baseline experiment, regardless of the estimated true effect size in the replications. This may be undesirable, especially in SE where experiments are commonly small \cite{dybaa2006systematic}. As Shepperd et al. put it \cite{shepperd2018role}, \textit{...so we may have the unintuitive situation that an underpowered original experiment showing a small effect may be a good deal easier to confirm than one might imagine!} This applies to experiments showing both a small and large effect, as it is the experiment sample sizes rather than the estimated effect size that drives the width of 95\% PIs \cite{patil2016should}.  

We do not think that this phenomenon can be addressed by setting subjective thresholds in SE (e.g., considering that a replication successfully reproduces the results of a baseline experiment if its effect size is within a specified range around the effect size of the baseline), because such thresholds may be dependent upon the technologies being evaluated. For example, one might claim that two replications, each providing a Cohen's d separated by just 0.1 (e.g., a Cohen's d of 0.2 in the baseline experiment, and a Cohen's d of 0.1 in the replication) successfully reproduce the results for inexpensive technologies (as, according to the rules of thumb, both effect sizes are small \cite{cohen1988statistical}). However, a 0.1 divergence in the results of a performance evaluation of expensive technologies may constitute a massive difference in terms of benefits. For example, investment in the \say{winning} technology may only be worthwhile if the effect size is 0.2 or larger, but not if the effect size is 0.1, as the replication suggested. In this case, the replication would have failed to reproduce the results of the baseline experiment. 

Besides, different stakeholders may set different thresholds. For example, while a difference of 0.1 between two replications may represent similar results for one developer (as the winning technology did not lead to a major improvement in either replication), this same difference of results may be relevant for a manager depending upon strict requirements for selecting a development approach for a team of developers. 

In summary, neither $p$-values, effect sizes, 95\% CIs, 95\% PIs, nor subjective thresholds appear to be suitable for assessing the extent to which the results of the replications hold. As a result of their use, there is a higher risk of the differences in results being attributed to contextual variables, when, in actual fact, they had no bearing at all on the contradictory findings.

\section{RQ1.2 Should Meta-Analysis be Used to Verify the Results of Previous Experiments?}
\label{proposal}

To gain an insight into the usefulness of meta-analysis, it may help to think about a hypothetical---and very large---experiment with 1,000 subjects per treatment (e.g., 1,000 subjects exercising ITL and 1,000 subjects exercising TDD). Again, suppose that, instead of calculating the results of the experiment as a whole, we calculated its results chunkwise. For example, this large-scale experiment could also also be conceived as an experiment composed of 10 identical chunks (or sub-experiments), each containing 100 subjects per treatment. The $p$-values, effect sizes, and 95\% CIs of these chunks would fluctuate around the total result of the large-scale experiment. However, if it had been divided into 100, instead of 10, chunks (with 10 subjects per treatment), the $p$-values, effect sizes and 95\% CIs of the respective chunks around the overall result would have been even less consistent with the overall outcome \cite{cumming2013understanding}. If we use meta-analysis to pool together the results of these chunks, the joint result will match up with the overall result of the large-scale experiment \cite{borenstein2011introduction}.

Figure~\ref{fig:chunks} illustrates this by showing the estimated effect sizes obtained across 5,000 simulated experiments of sample size 144 at different true effect sizes (i.e., true Cohen's d of 0.2, 0.5 and 0.8) when the effect sizes are estimated from of one single experiment, or meta-analyzing the results obtained in 4, 8 and 12 experiments---chunks---of sample sizes 36, 18 and 12 each. Since the four lines overlap, this can be better seen in Appendix~\ref{chunks}, where each line is shown in a separated graphic.

\begin{figure}[h!]  
  \caption{Joint result of meta-analysis vs. overall result of the large-scale experiment.}
  \label{fig:chunks}
  \centering
    \includegraphics[width=\textwidth,keepaspectratio]{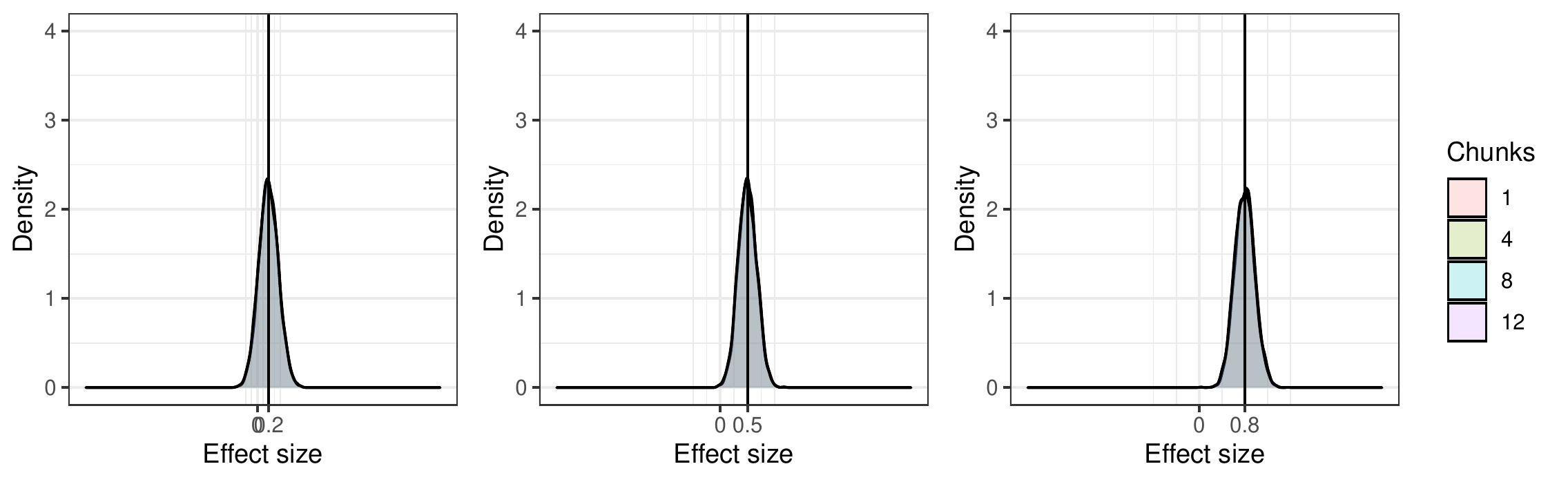}
\end{figure}

The relationship between the chunks and the large-scale experiment  and between replications and the population is exactly the same \cite{borenstein2011introduction}.  To illustrate this point, we now conduct a series of simulations. Each of them simulates 5,000 groups of replications with an identical sample size (estimating an identical true effect size) and an identical group size, whose results are then meta-analyzed to output the joint effect size. After we have simulated these 5,000 groups of replications, we pool their joint effect sizes into a distribution to show their variability. 

Figure \ref{meta_analysis} shows the distribution of joint effect sizes achieved in 5,000 groups of 1 to 12 identical AB between-subjects replications, each with a sample size of 36, meta-analyzed at different true effect sizes (i.e., Cohen's d of 0.2, 0.5 and 0.8, corresponding to a small, medium and large true effect size), from left to right, respectively \cite{cohen1988statistical}). A solid black vertical line indicates the estimated true effect size. Appendix~\ref{variability_effect_sizes} shows the results for sample sizes 4 and 100.

\begin{figure}[h!]  
  \caption{Meta-analysis: groups with different number of replications and true effect sizes.}
  \label{meta_analysis}
  \centering
    \includegraphics[width=\textwidth,keepaspectratio]{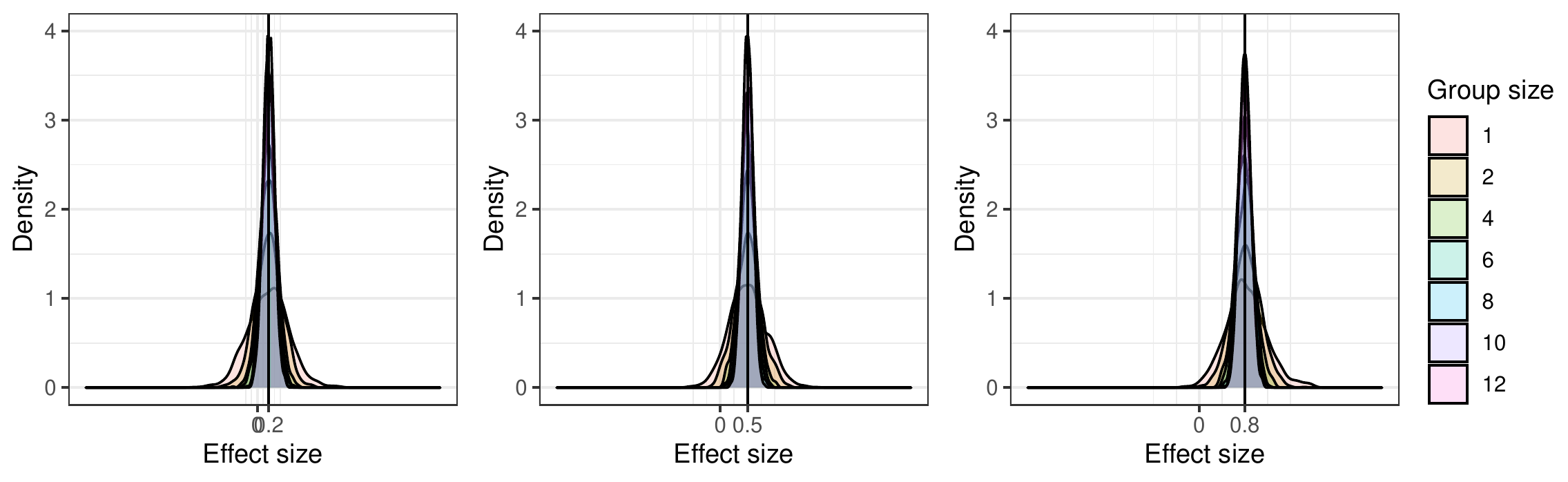}
\end{figure}

As Figure~\ref{meta_analysis} shows, the variability of the joint effect size decreases as the number of jointly meta-analyzed replications increases (the spread of the curves narrows as the number of meta-analyzed replications grows). Thus, even though each replication provides a different estimation of the true effect size (as a different sample participates in each replication), the joint estimated effect approaches the true effect size of the population when their estimated effect sizes are jointly meta-analyzed because all the replications estimate an identical true effect size. Table~\ref{confidence_intervals_2} shows the variation---95\% CIs---in the estimated effect sizes. Appendix~\ref{variability_effect_sizes} shows the results for sample sizes 4 and 100.

\begin{table}[!htbp] \centering 
  \caption{95\% confidence intervals for joint estimated effect sizes} 
  \label{confidence_intervals_2} 
\begin{tabular}{@{\extracolsep{5pt}} cccccc} 
\\[-1.8ex]\hline 
\hline \\[-1.8ex] 
True  & Group & \multicolumn{3}{c}{Joint estimated effect size} \\  \cline{3-5}
effect size & size & Lower bound 2.5\% & Mean & Upper bound 97.5\% \\ 
\hline \\[-1.8ex] 
$0.2$ & $1$ & $$-$0.48$ & $0.19$ & $0.90$ \\ 
& $2$ & $$-$0.27$ & $0.20$ & $0.66$ \\ 
& $4$ & $$-$0.11$ & $0.20$ & $0.52$ \\ 
& $6$ & $$-$0.08$ & $0.20$ & $0.46$ \\ 
& $8$ & $$-$0.02$ & $0.21$ & $0.43$ \\ 
& $10$ & $$-$0.01$ & $0.20$ & $0.42$ \\ 
& $12$ & $0.01$ & $0.20$ & $0.39$ \\ 
$0.5$ & $1$ & $$-$0.14$ & $0.51$ & $1.19$ \\ 
& $2$ & $0.03$ & $0.51$ & $1.03$ \\ 
& $4$ & $0.16$ & $0.51$ & $0.85$ \\ 
& $6$ & $0.23$ & $0.50$ & $0.79$ \\ 
& $8$ & $0.27$ & $0.51$ & $0.74$ \\ 
& $10$ & $0.29$ & $0.50$ & $0.71$ \\ 
& $12$ & $0.30$ & $0.50$ & $0.68$ \\ 
$0.8$ & $1$ & $0.16$ & $0.83$ & $1.62$ \\ 
& $2$ & $0.32$ & $0.82$ & $1.33$ \\ 
& $4$ & $0.47$ & $0.81$ & $1.16$ \\ 
& $6$ & $0.51$ & $0.80$ & $1.07$ \\ 
& $8$ & $0.57$ & $0.81$ & $1.06$ \\ 
& $10$ & $0.58$ & $0.80$ & $1.04$ \\ 
& $12$ & $0.60$ & $0.80$ & $1.00$ \\ 
\hline \\[-1.8ex] 
\end{tabular} 
\end{table}


In view of this, there is a straightforward argument in favor of the meta-analysis of replications in SE. According to Table~\ref{confidence_intervals}, a single replication of sample size 36 estimating a Cohen's d of 0.5 outputs a 95\% CI (-0.14,1.23). This means that if we replicate the experiment six times, we will get six effect sizes in that range. However, if we meta-analyze the six replications, the result will be one single joint effect size in the range (0.23,0.79). Finally, according to Figure~\ref{identical_results}, only the results of around 25\% of the replications (two at most) will be significant. Both will conclude that results are not reproducible, whereas the meta-analysis will find that they are.

In particular, as SE replications are usually small \cite{dybaa2006systematic} and, therefore, their effect sizes may be far removed from the estimated true effect size, the joint meta-analysis of the results of the replications may help to output more reliable results \cite{whitehead2002meta}\cite{borenstein2011introduction}\cite{button2013power}\cite{higgins2011cochrane}. 

Note that using meta-analysis instead or $p$-values or effect sizes gives us a slightly different view of the goal of replication. With meta-analysis, the results achieved in baseline experiments are no longer regarded as a result that needs to be reproduced, but as a small piece of evidence within a larger picture that only emerges after assembling (by means of meta-analysis) many small pieces (from each replication) to complete the puzzle. \\

\noindent\fbox{
  \parbox{\textwidth}{
    \textbf{Key findings}
    \begin{description}
 
    \item[\textbf{KF5}] Meta-analysis should be used to verify the results of previous experiments in SE. The estimated joint effect size obtained meta-analyzing several replications match the one that would have been estimated if the data from the replications would come from a single experiment.

    \item[\textbf{KF6}] The results of a SE baseline experiment should not be regarded as something to be reproduced, but as a small piece of evidence to be assembled with other pieces to provide a mature piece of knowledge.

    \end{description}
  }
}
\\

\section{RQ2 How Can We Discover if (Un)Intentional Contextual Modifications Have Altered the True Effect Size and Properly Interpret Results?}
\label{heterogeneity}

Some authors in SE (e.g., \cite{juristo2009using}) suggest that the failure of replications to reproduce the results of previous experiments in terms of $p$-values, effect size magnitudes (e.g., small, medium, or large \cite{borenstein2011introduction}) or signs (e.g., positive or negative) can be used to identify contextual variables that may have impacted the results (e.g., as the participants copied in the replication, this may have affected the results of the replication \cite{juristo2009using}). Unfortunately, this approach has a major shortcoming when $p$-values or estimated effect sizes are used to assess results reproducibility. As discussed in Section~\ref{checking}, experiments estimating the same true effect size---especially if they are small \cite{button2013power}\cite{cumming2013understanding}---may throw up conflicting results which are not necessarily being caused by any contextual variable. Thus, there is a risk of claiming that the difference between the results was caused by a contextual variable, when, in reality, this contextual variable may have played no role in the conflicting results.

We have seen, in Section~\ref{proposal},  that meta-analysis can help to deal with this problem, as it can be used to output joint results by pooling together replications estimating an identical true effect size. However, meta-analysis can also be used to pool together the results of replications estimating different true effect sizes. When running a meta-analysis, two types of models can be fitted: fixed-effects models or random-effects models \cite{borenstein2011introduction}. \textit{Fixed-effects models} assume that all the experiments are estimating a single true effect size. Thus, differences across the results of experiments may only emerge due to natural variation---that is, the different samples participating in each experiment. \textit{Random-effects models} assume that the experiments may be estimating different true effect sizes. Thus, differences across the results of experiments may emerge not only because of natural variation---as in fixed-effects models---but also due to the different true effect sizes being estimated in each experiment.

While fixed-effects models estimate a single joint effect size, the result of random-effects meta-analysis is a (usually normal) distribution of effect sizes \cite{borenstein2011introduction}. This distribution is centered at a joint effect size (that represents the mean effect size across the experiments), and has a variance equal to $\tau^2$. $\tau^2$ is a between-experiments variability parameter that represents the total variability of results that cannot be explained by natural variation \cite{borenstein2011introduction}. If $\tau^2$ is different from 0, then \textit{statistical heterogeneity} is said to have materialized \cite{borenstein2011introduction}. If statistical heterogeneity materializes, then each experiment may be estimating a potentially different true effect size. The larger $\tau^2$ is, the larger the statistical heterogeneity of results will be, and vice versa. 

The existence of heterogeneity can also be checked by means of the $Q$ statistic (and its associated $p$-value) and the $I^2$ statistic~\cite{borenstein2011introduction}. The $Q$ test addresses the homogeneity assumption---that is, tests the null hypothesis that all studies share a common effect size, and $p$ is its associated $p$-value. $I^2$  is used to determine what proportion of the observed variance is real. If it is near zero, then almost all observed variance is spurious, and there is nothing to explain. If it is large, then it would make sense to speculate about reasons for, and try to explain, the variance\footnote{I$^2$ is interpreted as: 25\% low, 50\% medium and 75\% high~\cite{higgins2003measuring}.}.

We recommend using $I^2$ to check for heterogeneity, as it is unaffected by three issues: first, $Q$ and its associated $p$-value depend strongly on the number of replications being meta-analyzed; second, the fact that $Q$ is not significant cannot be used as evidence for the existence of homogeneity; finally, $\tau^2$ depends on the metric used to measure effect size, which implies that it is meaningful to compare the values for two meta-analyses only when effect sizes are measured by the same metric.

In order to know the accuracy of $I^2$, its 95\% CI should be reported. An inaccurate value of $I^2$---wide CI--- means that more experiments are needed.

Researchers from other disciplines recommend the use of random-effects models rather than fixed-effects models \cite{borenstein2011introduction} \cite{whitehead2002meta}\cite{leandro2008meta} for several reasons: fixed-effects models may provide biased results if statistical heterogeneity materializes \cite{borenstein2011introduction}\cite{whitehead2002meta}, the results of random-effects and fixed-effects models are identical if statistical heterogeneity does not materialize (i.e., if $\tau^2$ is equal to 0) \cite{whitehead2002meta}\cite{borenstein2011introduction} and, if fixed-effects and random-effects models output different results, the results of random effects models tend to be more conservative---and thus, may err on the safe side \cite{whitehead2002meta}\cite{petitti2000meta}\cite{chen2013applied}\cite{lau1998summing}.

In view of this, and like others before us in other fields \cite{borenstein2011introduction}\cite{whitehead2002meta}, we recommend the use of random-effects meta-analysis models to pool together the results of SE replications. Our recommendation is based on the complexity of the SE context, where many variables may lead to statistical heterogeneity in SE\footnote{Note, however, that there are other alternatives to random-effects meta-analysis. Empirical Bayes, for instance, has the advantage of being more explicit and using a better approximation algorithm.}.

If statistical heterogeneity materializes, researchers should strive to identify potential sources of heterogeneity across the experiments \cite{borenstein2011introduction}\cite{whitehead2002meta}. According to mature experimental disciplines, such as medicine \cite{gagnier2012investigating}\cite{higgins2011cochrane}, statistical heterogeneity may emerge as a consequence of two different sources of heterogeneity. The first one is due to different experimental configurations of the experiments \cite{higgins2011cochrane}. In SE, this has to do with the different experimental designs, programming languages, session lengths, etc., of the experiments. Hereinafter we will refer to this question as \textit{configuration heterogeneity}. The second one is due to different characteristics of the participants across the experiments \cite{higgins2011cochrane}. In SE, it is related to the different skills, backgrounds, experiences, etc., of the participants across the experiments. Hereinafter we will refer to this question as \textit{sample heterogeneity}.}

Once potential sources of heterogeneity (configuration or sample variables)  have been identified, we can apply techniques, such as linear mixed models, subgroup analyses or meta-regression, to check their potential impact on results.

However, failure to find statistical heterogeneity at experiment level does not imply that heterogeneity did not materialize at participant level \cite{simmonds2005meta}\cite{groenwold2010subgroup}. In particular, suppose we have replications with similar results whose participants---of potentially different types---achieve different results. Say, for example, that a similar number of novice developers and senior developers participate in all the replications, and, thus, the average results of the replications are similar, but the novice and senior developers actually perform differently with the technologies under evaluation. Thus, regardless of whether or not statistical heterogeneity materializes across the replications, the extent to which participant characteristics impacts the results should be assessed in all cases \cite{simmonds2005meta}\cite{groenwold2010subgroup}, provided that the raw data are accessible and the characteristics of the participants have been measured.

Although statistical techniques such as sub-group meta-analysis or linear mixed models are key for assessing the effect of changes made across replications \cite{borenstein2011introduction}\cite{cumming2013understanding}\cite{brown2014applied}\cite{fisher2011critical}, they are not a panacea. Even though these statistical techniques indicate that the changes made across the replications may have affected the results, the difference in the results may have also been caused by other variables (i.e., confounding variables) rather than the changes made across the replications \cite{lau1998summing}\cite{cumming2013understanding}\cite{borenstein2011introduction}\cite{adriguidelines}. Suppose, for example, we run one replication with professionals and another with students, any statistically significant difference between the results of both replications identified by the sub-group meta-analysis may potentially have been caused by variations in the age of the participants (as professionals tend to be older than students), their motivation (as professionals may be less motivated than students), or their treatment conformance (as professionals may have had to attend to their daily work activities, which may have prevented them from adhering closely to the treatment application procedure) and so on rather than to the different types of participants across the replications. Thus, such analyses are exploratory and may, at best, serve for guiding prospective studies---rather than for providing definitive answers \cite{lau1998summing}\cite{cumming2013understanding}\cite{borenstein2011introduction}\cite{adriguidelines}.\\

\noindent\fbox{
  \parbox{\textwidth}{
    \textbf{Key findings.} When meta-analyzing the results of SE replications: 
    \begin{description}

    \item[\textbf{KF7}] Random-effects models should always be used, checking, in all cases, for heterogeneity due to possible inadvertent contextual changes.

    \item[\textbf{KF8}] If heterogeneity does materialize, configuration variables should be identified, assessing their impact using subgroup analyses or meta-regression, for example.

    \item[\textbf{KF9}] Regardless of the existence of heterogeneity, it is important to always check for sample heterogeneity using linear mixed models, for example.

    \item[\textbf{KF10}] When the statistical heterogeneity value is inaccurate, more replications need to be run.

	\item[\textbf{KF11}] Access to the raw data is required to perform sample heterogeneity meta-analysis.

    \item[\textbf{KF12}] The statistical techniques used for assessing the effect of changes made across replications have limitations.
       
    \end{description}
  }
}
\\

\section{Examples}
\label{examples}

In the following, we show three real-life groups of different replications from our own research. In particular, Section \ref{example_1} reports replications with identical configurations and sample characteristics (at the same university, over successive academic years), where we tried to remove statistical heterogeneity by design. Then, Section \ref{example_2} reports replications with different sample characteristics (run across different sites within the same multinational company), where we accidentally introduced sample heterogeneity that eventually led to statistical heterogeneity. Finally, Section \ref{example_3} reports replications with different configurations (different technological environments and trainers), where we introduced configuration heterogeneity that eventually led to statistical heterogeneity.

\subsection{Replications with Identical Configurations and Sample Characteristics}
\label{example_1}

Some years ago \cite{juristo2012comparing}, we ran a series of identical replications---over successive academic years---at a Spanish university (UPM) to compare the effectiveness of two testing techniques: equivalence class partitioning (EP), and branch testing (BT). We measured the effectiveness of each testing technique as the percentage of faults that the participants detected when applying such techniques to test one out of three different toy programs (i.e., cmdline, ntree, nametbl). We counterbalanced the programs to the testing techniques to rule out the influence of the program on results. We ran a total of five exact replications between 2001 and 2005. A total of 35, 39, 29, 36 and 27 subjects participated in UPM01, UPM02, UPM03, UPM04, UPM05, respectively\footnote{For a detailed description of the experiments, their designs, and results please refer to \cite{juristo2012comparing}.}. 

\begin{figure}[h!]  
  \caption{Forest plot: EP vs. BT.}
  \label{forest_plot_example_1}
  \centering
    \includegraphics[width=\textwidth,keepaspectratio]{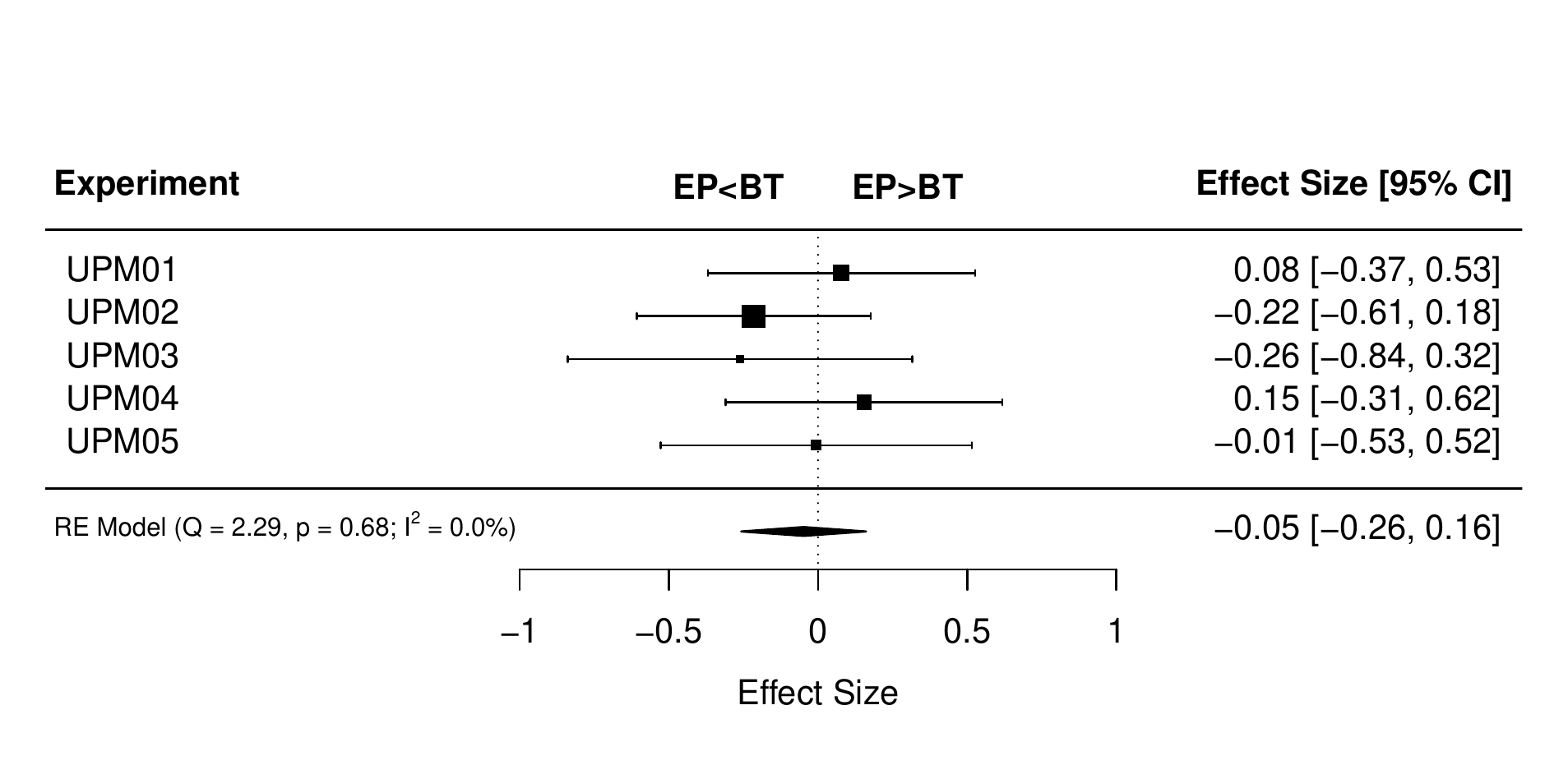}
\end{figure}

We analyzed each replication separately. Table~\ref{ttest_example1} shows the results of the t-test. Additionally, we calculated the Cohen's d effect size for the testing techniques in each replication, which were later pooled together by means of random-effects meta-analysis \cite{borenstein2011introduction} (following KF1, KF3, KF5, KF6 and KF7). Figure~\ref{forest_plot_example_1} shows the forest-plot of this meta-analysis.

\begin{table}[!h] \centering 
  \caption{Results of the t-tests.} 
  \label{ttest_example1} 
\begin{tabular}{lrrr} \hline \hline
\textbf{Experiment} & \textbf{t} & \textbf{df} & \textbf{$p$-value} \\ \hline
UPM01 &  -0.34 & 34 & 0.73 \\
UPM02 & 1.09 & 38 & 0.28 \\
UPM03 &  0.91 & 28 & 0.37 \\
UPM04 &  -0.65 & 35 & 0.52 \\ 
UPM05 &  -0.02 & 26 & 0.98 \\ \hline
\end{tabular} 
\end{table}

On the one hand, Table~\ref{ttest_example1} shows that the null hypothesis of equality of means cannot be rejected in any of the experiments. On the other hand, Figure \ref{forest_plot_example_1} shows that, even though the replications were identical, they provided seemingly conflicting effect sizes (i.e., either EP$>$BT or EP$<$BT). Notice that despite the wide 95\% CIs that materialized in the experiments (see the wide lines that cross the effect size of each experiment), the 95\% CI of the joint effect size (i.e., the black diamond at the bottom of the forest-plot) is much narrower, as expected. Thus, although the replications provided conflicting estimated effect sizes (leading us to think that the results were not being reproduced, as they were both positive or negative \cite{juristo2009using}), the meta-analysis provided a much more precise joint effect size (in line with KF2 and KF4). 

Also, looking at the 95\% CIs of the experiments, we find that they overlap with each other. This suggests that no statistical heterogeneity of results materialized---as the effect sizes of the replications fell within the expected bounds of the 95\% CIs of the others \cite{cumming2013understanding}\cite{borenstein2011introduction}. The $I^2$ statistic appears to corroborate the statistical homogeneity statistic: as $I^2=0$. This indicates that $\tau^2=0$, and, thus, according to meta-analysis, 0\% of the total variation is due to statistical heterogeneity  \cite{borenstein2011introduction}. However, the 95\% CI for $I^2$ is (0.0\%,78.5\%).

Unfortunately (following KF8 and KF9), we cannot identify any characteristics of the experimental configurations of the replications that may be behind the heterogeneity of results at UPM---as, at least in principle, all the replications had an identical experimental configuration, and all participants belonged to the same population. Participants can be considered different samples of the same population, as the syllabus of the degree did not change over the five years in question, and the students (undergraduates) all had the same background and profile.

If we assume that no statistical heterogeneity materialized, we could say that meta-analysis points towards the existence of a single true effect size in the population (to which this sample belongs\footnote{A population of novice testers with limited experience in software development, 12 hours of training on testing techniques,  testing toy programs. }) corresponding to a Cohen's $d=-0.05,~95\%~CI=(-0.26, 0.16)$. Thus, the results of the meta-analysis suggest that the effect size in the population (i.e., the true effect size being estimated by all the replications) is low \cite{higgins2003measuring}---as the 95\% CI of the joint effect size indicates that the true effect size of the populations can be expected to be within the range of -0.26 to 0.16. However, the true effect size could even be 0 (suggesting that the testing techniques perform identically in the population), as 0 is within the 95\% CI. If we want to achieve a greater precision of results, then we would need to run more replications, which would have to be meta-analyzed.

On the other hand, statistical heterogeneity could have materialized, as the resulting value of $I^2$ provided is inaccurate, due to the small number of replications within the meta-analysis. Therefore, more replications are required to clarify whether or not heterogeneity materializes (as indicated by KF10).

\subsection{Replications with Different Sample Characteristics}
\label{example_2}

Recently \cite{tosun2017industry}, we ran a series of identical replications at F-Secure (i.e., a multinational cybersecurity solutions company) to compare the performance of TDD and ITL in terms of external quality. We ran a total of three exact replications, each at a different location: Helsinki, Kuala-Lumpur and Oulu (i.e., F-Secure H, K and O, respectively). A total of 6, 11 and 7 professionals participated in each experiment, respectively\footnote{For a detailed description of the experiments, their designs, and results please refer to \cite{tosun2017industry}.}.

\begin{figure}[h!]  
  \caption{Mean experience of the participants across replications.}
  \label{experience_plot_example_2}
  \centering
    \includegraphics[width=9cm,keepaspectratio]{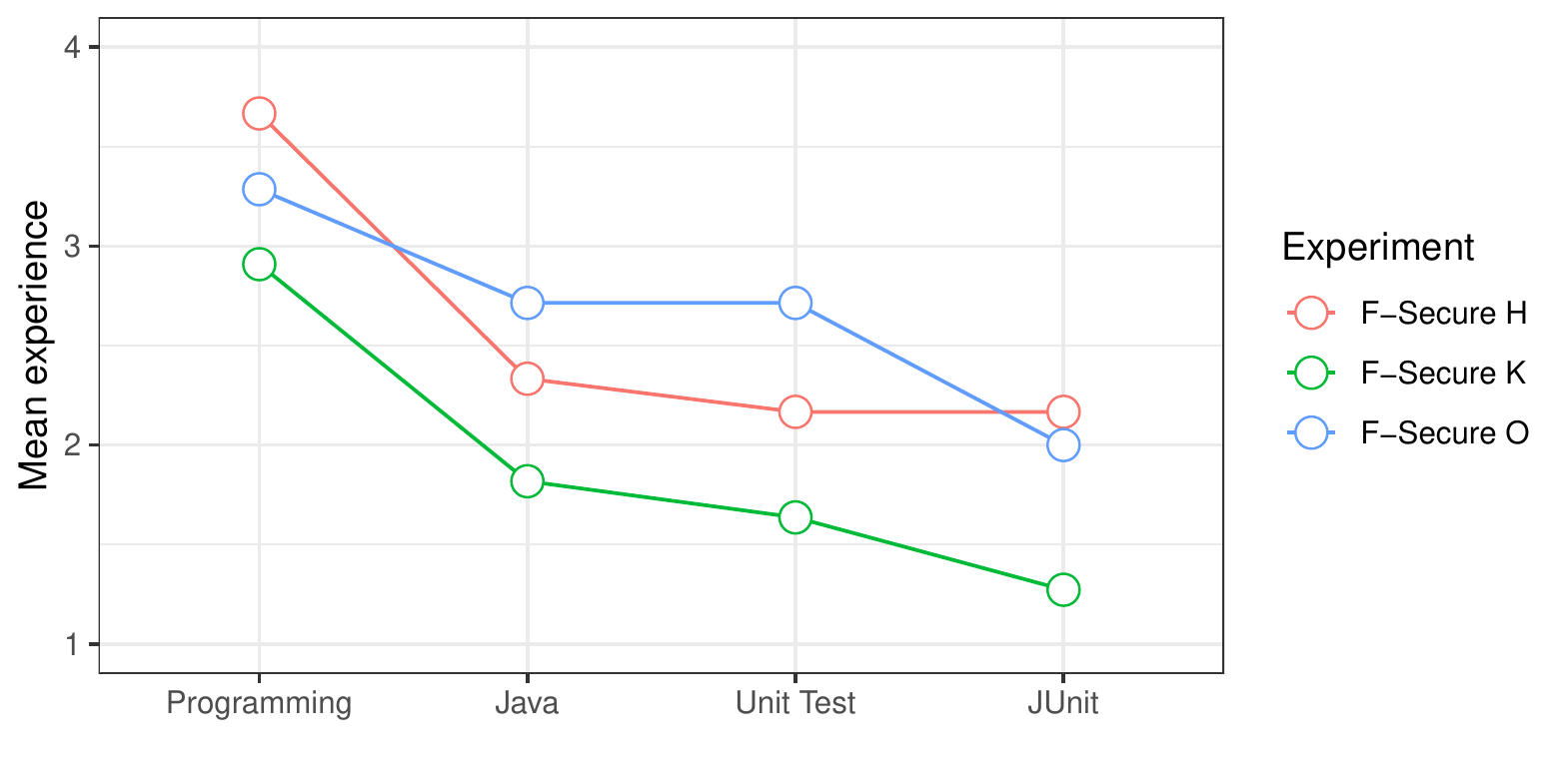}
\end{figure}

We handed out a survey to the participants a few days before the experiments were run. The survey contained a series of ordinal-scale (i.e., inexperienced, novice, intermediate and expert) self-assessment questions asking about their experience with programming, unit testing, Java and JUnit\footnote{The survey and its results were published elsewhere \cite{dieste2017empirical}.}. Figure \ref{experience_plot_example_2} shows the profile plot corresponding to the mean of the experiences of the participants across the replications (1-4, for inexperienced, novice, intermediate and experts, respectively)\footnote{For simplicity's sake, we consider the variables measured throughout the survey as continuous as in \cite{dieste2017empirical}.}. 

\begin{table}[!h] \centering 
  \caption{Summary statistics for participants' experience} 
  \label{experience_descriptive_statistics} 
\begin{tabular}{@{\extracolsep{5pt}} lccccc} 
\\[-1.8ex]\hline 
\hline \\[-1.8ex] 
Experience & Experiment & Mean & Median & Min. & Max. \\ 
\hline \\[-1.8ex] 
Programming & F-Secure H & $3.67$ & $4$ & $3$ & $4$ \\ 
& F-Secure K & $2.91$ & $3$ & $2$ & $4$ \\ 
& F-Secure O & $3.29$ & $3$ & $2$ & $4$ \\ 
Java & F-Secure H & $2.33$ & $2.50$ & $1$ & $4$ \\ 
& F-Secure K & $1.82$ & $2$ & $1$ & $4$ \\ 
& F-Secure O & $2.71$ & $3$ & $1$ & $4$ \\ 
Unit Testing & F-Secure H & $2.17$ & $2$ & $1$ & $4$ \\ 
& F-Secure K & $1.64$ & $2$ & $1$ & $2$ \\ 
& F-Secure O & $2.71$ & $3$ & $2$ & $4$ \\ 
JUnit & F-Secure H & $2.17$ & $2$ & $1$ & $4$ \\ 
& F-Secure K & $1.27$ & $1$ & $1$ & $2$ \\ 
& F-Secure O & $2$ & $2$ & $1$ & $3$ \\ 
\hline \\[-1.8ex] 
\end{tabular} 
\end{table}

As Figure~\ref{experience_plot_example_2} shows, a heterogeneous population of developers participated in the replications: ranging from the most senior to the most novice developers at F-Secure H and O and at F-Secure K, respectively. According to the minimum and maximum values in Table~\ref{experience_descriptive_statistics}, there seems to be heterogeneity at participant level within each experiment as well.

\begin{figure}[h!]  
  \caption{Forest plot: TDD vs. ITL.}
  \label{forest_plot_example_2}
  \centering
    \includegraphics[width=\textwidth,keepaspectratio]{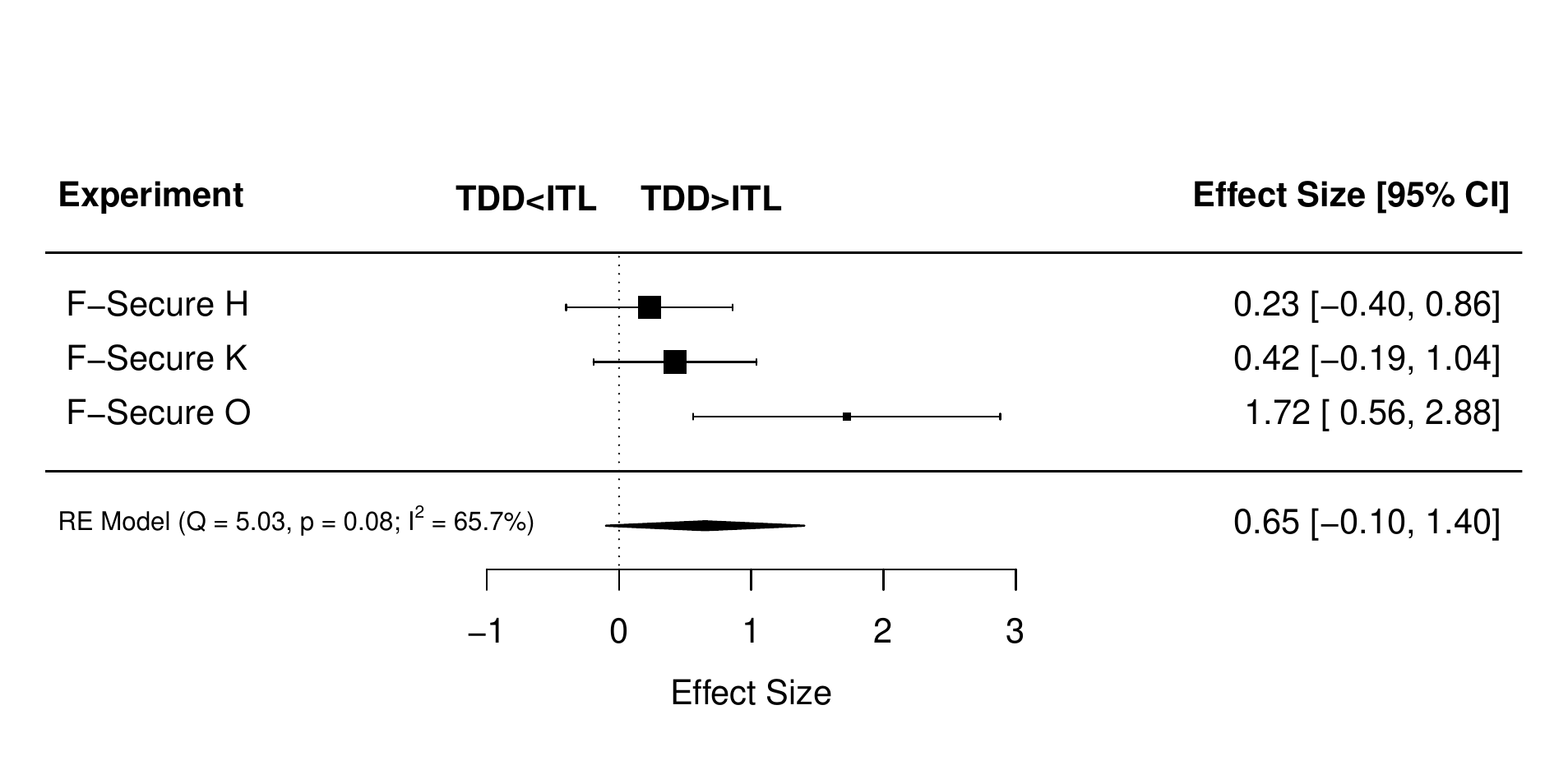}
\end{figure}

We analyzed each replication separately. Table~\ref{ttest_example2} shows the results of the t-test. Additionally, we calculated the Cohen's d effect size of each replication (i.e., the difference between the mean performance achieved with TDD and ITL in standardized units), which were then pooled together by means of random-effects meta-analysis \cite{borenstein2011introduction} (following KF1, KF3, KF5, KF6 and KF7). Figure~\ref{forest_plot_example_2} shows the forest-plot of this meta-analysis.

\begin{table}[!h] 
  \centering 
  \caption{Results of the t-tests.} 
  \label{ttest_example2} 
\begin{tabular}{lrrr} \hline \hline
\textbf{Experiment} & \textbf{t} & \textbf{df} & \textbf{$p$-value} \\ \hline
F-Secure H & -0.73 & 5 & 0.50 \\
F-Secure K & -1.35 & 10 & 0.21 \\ 
F-Secure O & -5.13 & 6 & 0.00 \\ \hline
\end{tabular} 
\end{table}

Table~\ref{ttest_example2} shows contradictory results. While the null hypothesis of equality of means could be rejected in one of the experiments (F-Secure O), this was not possible in the other two (F-Secure H and K). Additionally, Figure \ref{forest_plot_example_2} shows that, even though TDD outperforms ITL in all three replications (the effect sizes of all the replications are placed on the right hand-side of the forest-plot), the extent of the outperformance is largely dependent upon the site where the experiment was run. In particular, TDD outperforms ITL to a much larger extent at F-Secure O than in the other experiments. Surprisingly, even though the replications were identical (insofar as they were conducted by the same researchers, using identical experimental configurations, and with professionals from the same company), the statistical heterogeneity that materialized when the results were combined using meta-analysis (i.e., $I^2=65.7\%$) was rather large---medium according to the rules of thumb \cite{borenstein2011introduction}. However, the 95\% CI for $I^2$ is (0.0\%,99.4\%), meaning the obtained value is inaccurate and heterogeneity might not have materialized.

Thus, according to meta-analysis, the replications might be estimating different true effect sizes. If this is the case, rather than providing a single true effect size with meta-analysis (as we did in the previous group of replications), the best that meta-analysis can do here is to provide a distribution of true effect sizes in the population \cite{borenstein2011introduction}\cite{button2013power}. 

In particular, according to meta-analysis, a normal distribution of effect sizes with a mean equal to $M=0.65$ and a between-study variance equal to $\tau^2=0.28$\footnote{This estimate was calculated based on the output of the meta-analysis that we undertook with the \texttt{metafor} R package \cite{viechtbauer2010metafor}.}$^,$\footnote{$\tau^2$ can be estimated with different estimation methods, each of which may provide a potentially different estimate \cite{langan2018comparison}. In this article, we use restricted maximum likelihood (REML),  recommended by Langan et al., applicable to continuous outcomes \cite{langan2018comparison}. A large number of experiments are needed to estimate precise $\tau^2$ parameters (i.e., 5 \cite{feaster2011modeling}, 10 \cite{snijders2011multilevel}, 15 or even more \cite{mcneish2016effect}).} may be present in the population (i.e., a $\mathcal{N}(0.65,0.28)$ distribution of true effect sizes). Unfortunately, such distribution may be inaccurate due to the small number of replications that we meta-analyzed together \cite{whitehead2002meta}\cite{borenstein2011introduction}. Thus, more replications are still needed to increase the precision of results \cite{whitehead2002meta}\cite{borenstein2011introduction} (according to KF10).

Finally, as statistical heterogeneity might have materialized with meta-analysis, we should strive to identify potential sources of heterogeneity across the experiments \cite{borenstein2011introduction}\cite{whitehead2002meta} (in line with KF8 and KF9). In particular, and as previously mentioned in Section \ref{heterogeneity}, statistical heterogeneity may have emerged either because of the different experimental configurations of the replications (i.e., configuration heterogeneity) or because of the different characteristics of the subjects participating across the experiments (i.e., sample heterogeneity). 

Unfortunately, we cannot identify any of the characteristics of the experimental configurations of the replications that may be behind the heterogeneity of results at F-Secure---as, at least in principle, all the replications had an identical experimental configuration. However, the participants did have different levels of programming, unit testing, Java and JUnit experience, which may be affecting the results of the experiments. As we have access to the raw data, we can investigate the extent to which the experience of the participants may be impacting the results using linear mixed models (LMMs) with interaction terms \cite{brown2014applied}\cite{fisher2011critical} (as required by KF11). 

We applied Fisher et al.'s approach \cite{fisher2011critical} to conduct this analysis. In particular, we account for the heterogeneity of results by considering the development approach and the participant as random effects across the experiments. To assess the influence of participant characteristics on the results, we fit four different linear mixed models (one for each of the four characteristics), with the main effects of the development approach and the respective characteristic, their interaction, and the interaction between the development approach and the mean experience in each experiment. Table \ref{results_moderators} shows the results of the LMMs that we fitted to assess the influence of participant characteristics.

\begin{table}[!h] \centering 
  \caption{Effects of the characteristics of the participants on results} 
  \label{results_moderators} 
\begin{tabular}{lcccc} \hline \hline
\textbf{Experience} & \textbf{Estimate} & \textbf{$t$-value} & \textbf{$p$-value} \\ \hline
Programming & 16.79  & 1.83& 0.08 \\
Java & 3.45  &  0.52 & 0.60 \\
Unit Testing & 6.87  & 0.74 & 0.46 \\
JUnit & 8.03  & 0.96 & 0.34 \\ \hline
\end{tabular} 
\end{table}

\begin{figure}[h!]  
  \caption{Participant-level moderators on TDD.}
  \label{moderators_regression_line}
  \centering
    \includegraphics[width=9cm,keepaspectratio]{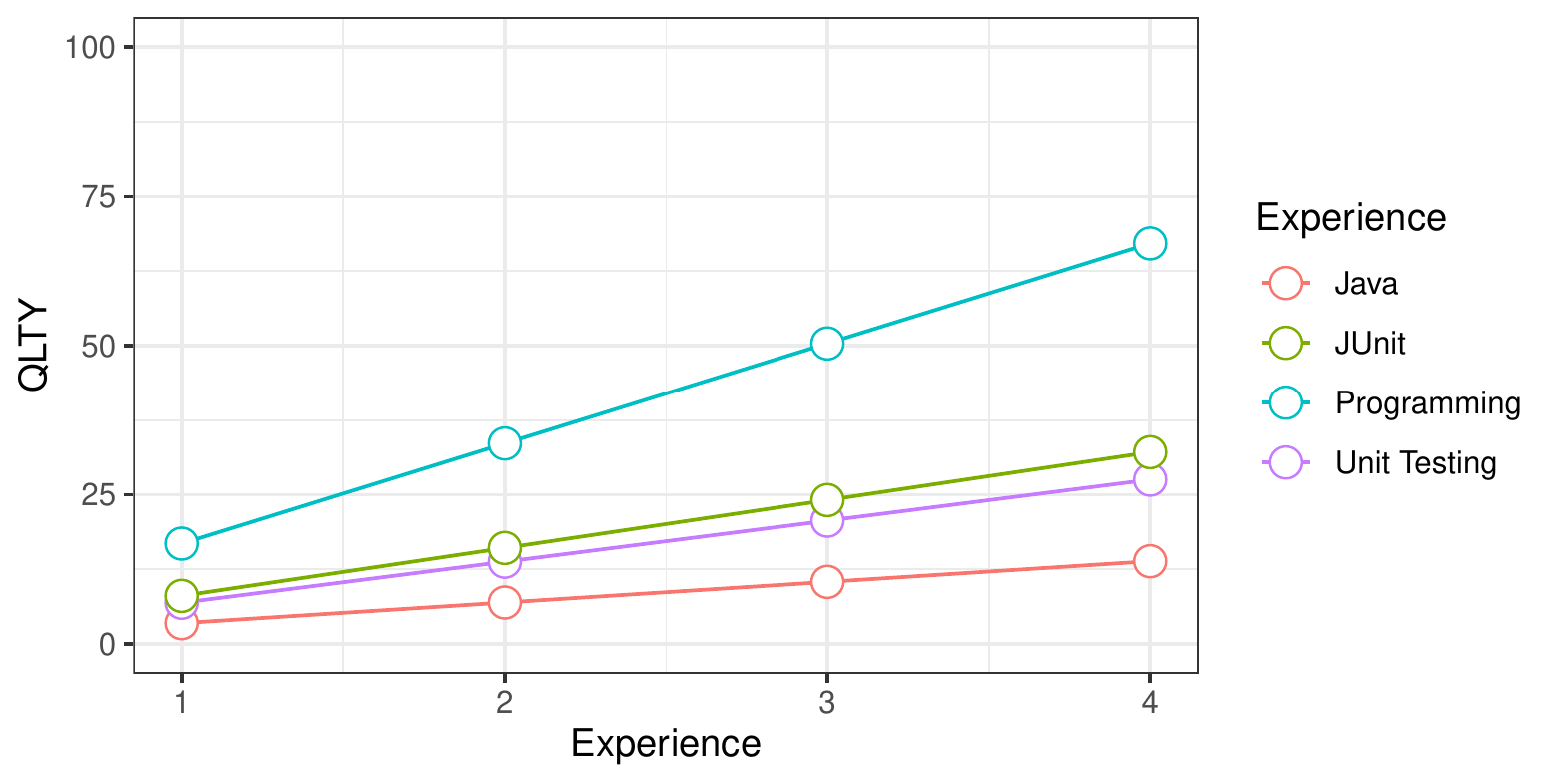}
\end{figure}

Table~\ref{results_moderators} can be read as the effect of a one-unit increase in experience on the performance achieved with TDD beyond the performance achieved with ITL. For example, for every unit increase in programming experience (see Table \ref{results_moderators}, Programming), performance with TDD increases by $M=16.79$ units. Thus, the more programming experience developers have (ranging from 1 to 4), the more TDD outperforms ITL. With the aim of easing the understanding of results, Figure \ref{moderators_regression_line} shows the regression lines for the results shown in Table \ref{results_moderators}.

As Figure \ref{moderators_regression_line} shows, the more experience participants have generally (with programming, Java, unit testing or JUnit), the more TDD outperforms ITL (notice that all the lines have an upward trend). In addition, programming experience appears to be the experience variable that has a bigger impact on results (as it is the line with the steepest slope). However, more experiments are needed to confirm these results (according to KF12).

\subsection{Replications with Different Configurations}
\label{example_3}

We ran a series of replications of the experiments that we ran at F-Secure at three different locations: UPM (a Spanish University), PlayTech (a multinational online gaming software supplier), and Ericsson (a multinational telecommunications company). A total of 14, 16, and 20 subjects participated in each experiment, respectively. We adapted the programming environments (i.e., the programming language and the testing tools) of the experiments to the expertise of the participants at each site. Also, at PlayTech, a trainer with lower experience was used. Table \ref{tab:experimental_design} shows the sample sizes, the programming environments and trainer experience of the replications that we ran.

\begin{table}[h!]
\small
\begin{center}
\caption{Characteristics of the replications on TDD.}
\label{tab:experimental_design}
\begin{tabular}{ l| l | l | l} \hline \hline
\textbf{Experiment} & \textbf{N} & \textbf{Environment} & \textbf{Trainer experience} \\ \hline
UPM & 14 & Java, JUnit & High \\
PlayTech & 16 & Java, JUnit & Low \\
Ericsson & 20 & C++, Boost  & High \\  \hline
\end{tabular}
\end{center}
\end{table}

As Table \ref{tab:experimental_design} shows, UPM and PlayTech used identical technological environments (i.e., Java and JUnit), unlike Ericsson (i.e., C++ and Boost). On the other hand, the trainer experience at Ericsson and UPM was the same, higher than the one at PlayTech.

\begin{figure}[h!]  
  \caption{Forest plot: TDD vs. ITL.}
  \label{experience_plot_example_3}
  \centering
    \includegraphics[width=\textwidth,keepaspectratio]{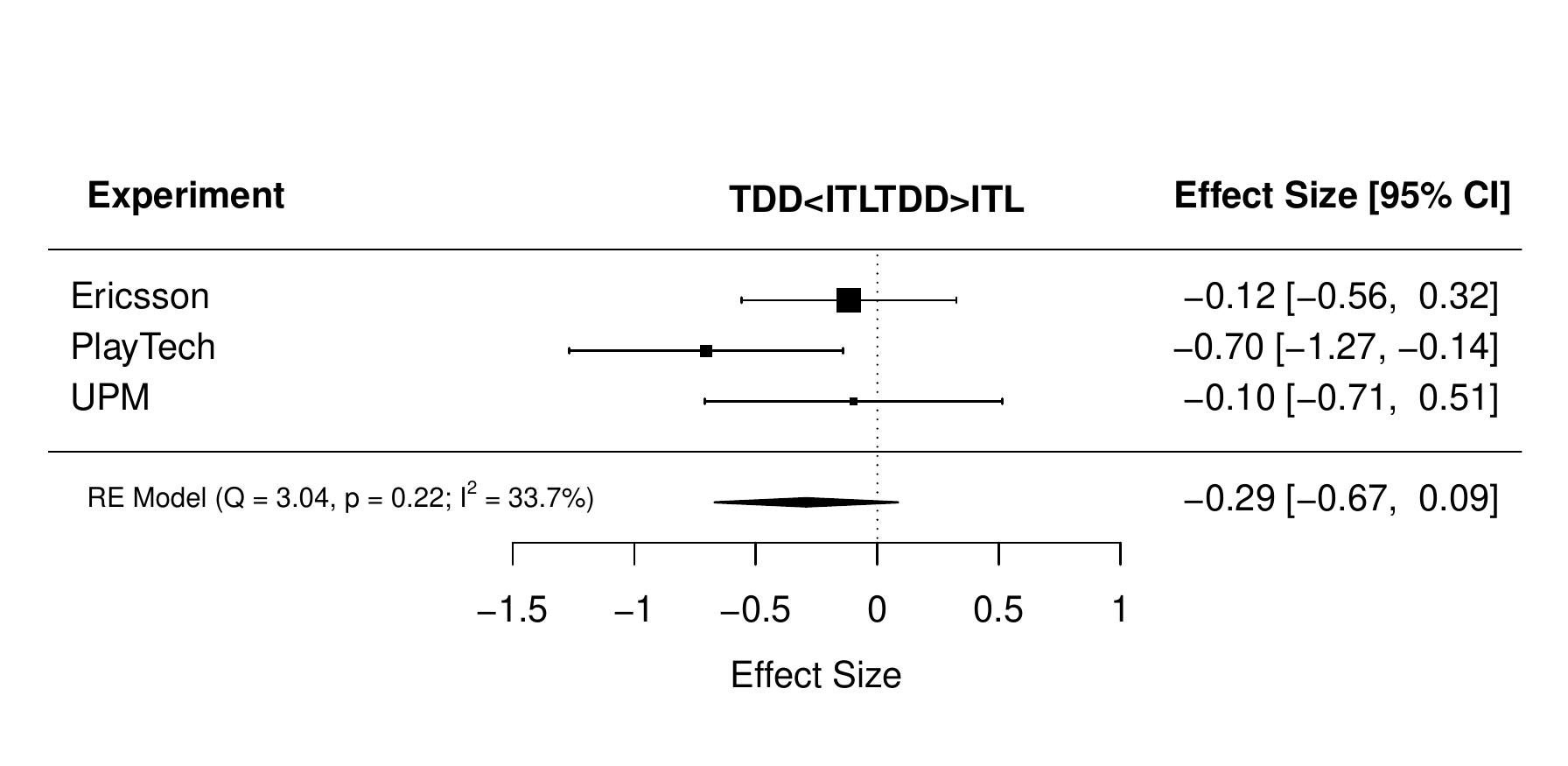}
\end{figure}

We analyzed each replication separately. Table~\ref{ttest_example3} shows the results of the t-test. Additionally, we calculated the Cohen's d effect size of each replication (i.e., the difference between the mean performance achieved with TDD and ITL in standardized units), which were then pooled together by means of random-effects meta-analysis \cite{borenstein2011introduction} (following KF1, KF3, KF5, KF6 and KF7). Figure \ref{experience_plot_example_3} shows the forest-plot for this meta-analysis. 

\begin{table}[!h] \centering 
  \caption{Results of the t-tests.} 
  \label{ttest_example3} 
\begin{tabular}{lrrr} \hline \hline
\textbf{Experiment} & \textbf{Estimate} & \textbf{$t$-value} & \textbf{$p$-value} \\ \hline
Ericsson & 0.51 & 19 & 0.62 \\
PlayTech & 2.73 & 15 & 0.02 \\
UPM & 0.31 & 13 & 0.76 \\ \hline
\end{tabular} 
\end{table}

Table~\ref{ttest_example3} shows contradictory results. While the null hypothesis of equality of means can be rejected in one of the experiments (PlayTech), this is not possible in the other two (as predicted by KF2). Additionally, as Figure~\ref{experience_plot_example_3} shows, ITL outperforms TDD across all the replications, albeit to a different extent: while ITL outperforms TDD by a long way at PlayTech, the difference in performance between TDD and ITL is much smaller at Ericsson and UPM (as predicted by KF4). In addition, a medium level of heterogeneity of results materialized, as $I^2=33.7\%$. However, note that the 95\% CI for $I^2$ is (0.0\%,98.41\%), meaning the obtained value is inaccurate and heterogeneity might not have materialized. Thus, according to the meta-analysis, the replications may be estimating different true effect sizes. Accordingly, the joint result of meta-analysis merely outputs a normal distribution of effect sizes centered at $M=-0.26$ and a variance equal to $\tau^2=0.047$ (i.e., the distribution of effect sizes in the population follows a $\mathcal{N}$(-0.26,0.047)). Unfortunately, this distribution of effect sizes may be inaccurate due to the small number of replications that we meta-analyzed together \cite{borenstein2011introduction}. Thus, more replications are still needed to increase the precision of results \cite{whitehead2002meta}\cite{borenstein2011introduction}, and clarify the existence of heterogeneity (according to KF10).

\begin{figure}[h!]  
  \caption{Sub-group meta-analysis: technological environment.}
  \label{experience_plot_example_3_subgroup1}
  \centering
    \includegraphics[width=\textwidth,keepaspectratio]{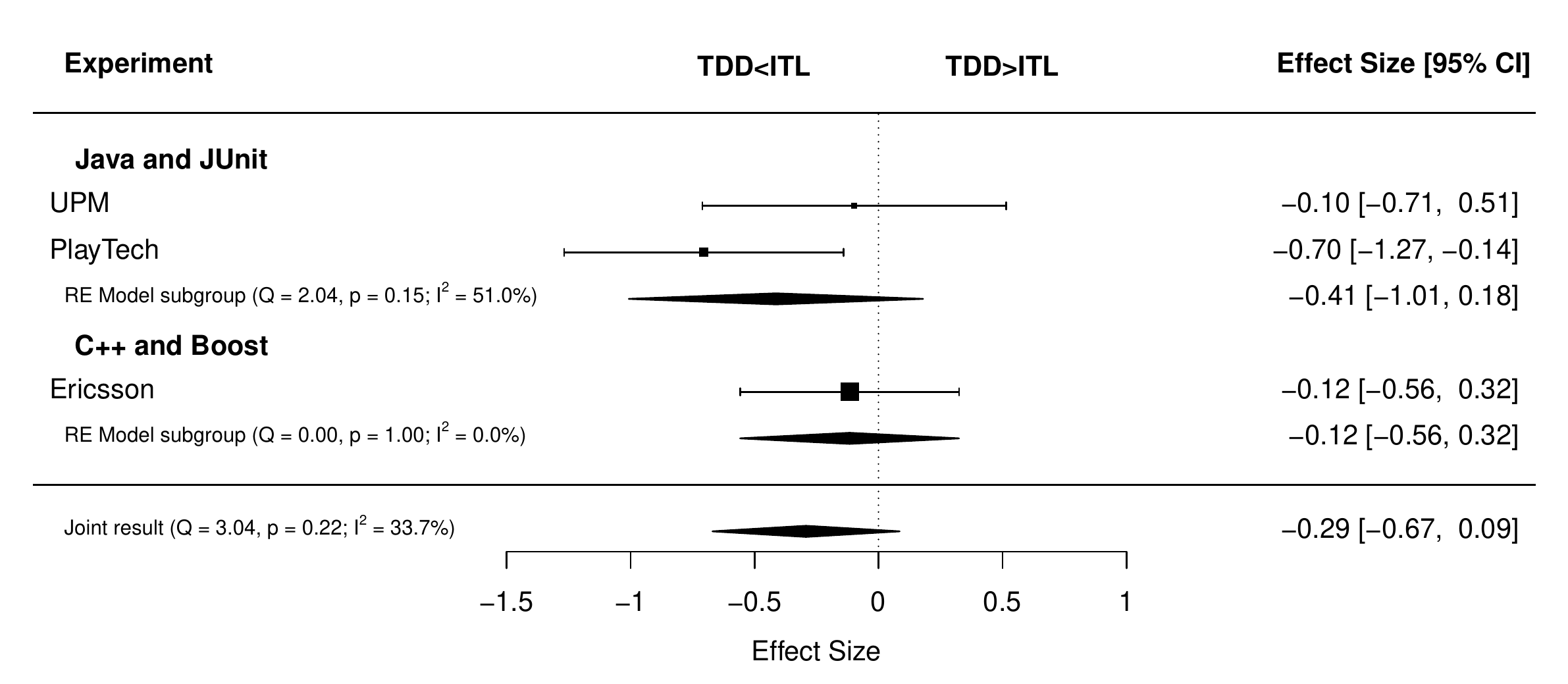}

\end{figure}

As the replications have different experimental configurations, this may be influencing their results. We can assess the extent to which the different technological environments used across the replications may have affected their results by means of sub-group meta-analysis \cite{borenstein2011introduction} (as suggested by KF8). Figure \ref{experience_plot_example_3_subgroup1} shows the forest-plot for the sub-group meta-analysis that we performed to assess this issue.

As Figure \ref{experience_plot_example_3_subgroup1} shows, although the sub-group composed of UPM and PlayTech had an identical programming environment, a medium level of heterogeneity of results still materialized ($I2=51.0\%$). But again, the obtained value is inaccurate (0.00\%-99.91\%). If we assume that heterogeneity has taken place, this may suggest that as both replications had an identical programming environment, other characteristics of the replications, or of the participants, may be behind their different results. Finally, the joint effect sizes for both programming environments (i.e., the joint effect size for Java and JUnit and the joint effect size for C++ and Boost) largely overlapped. This suggests that, although the replications had different technological environments, the technological environment may not be able to explain the difference of results achieved across the replications. 

\begin{figure}[h!]  
  \caption{Sub-group meta-analysis: training.}
  \label{experience_plot_example_3_subgroup2}
  \centering
    \includegraphics[width=\textwidth,keepaspectratio]{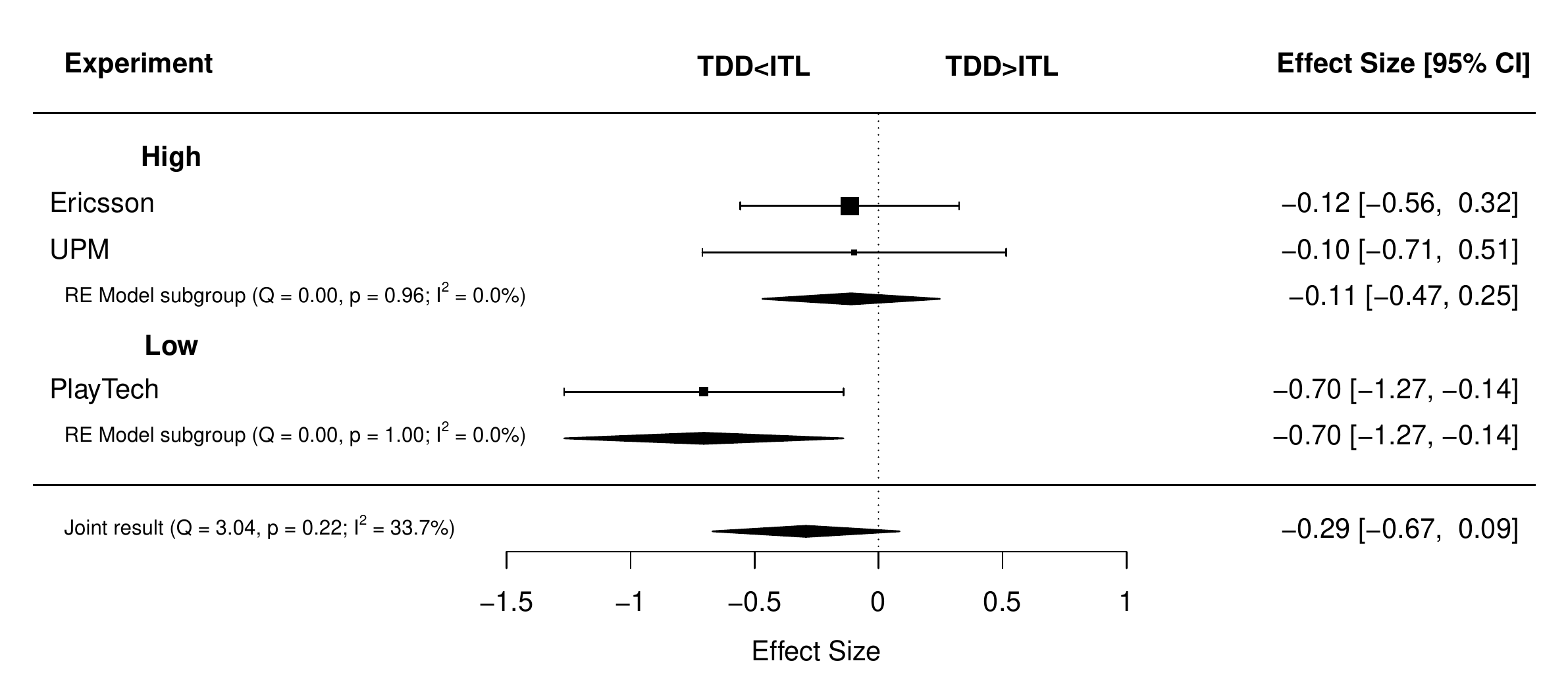}
\end{figure}

In view of the results, we also assessed the extent to which the experience of the trainers may have affected the results of the replications. Figure \ref{experience_plot_example_3_subgroup2} shows the forest-plot for the sub-group meta-analysis that we performed to assess this issue.

As Figure \ref{experience_plot_example_3_subgroup2} shows, no heterogeneity of results materialized in the sub-group composed of UPM and Ericsson ($I^2=0.0\%$). But again, the obtained value is inaccurate (0.00\%-58.34\%). This suggests that training could be behind the different results, although more experiments are needed to confirm these results (according to KF12).

\subsection{Discussion}
\label{discussion}

Our first example revealed conflicting results. However, neither contradictory $p$-values nor a positive and negative effect size in identical replications (e.g., a Cohen's d of 0.15 and -0.26, respectively) imply that unacknowledged characteristics of either the experiments or the participants impacted the results \cite{juristo2009using}.  Indeed, sampling error, aggravated by small sample sizes, may be at play, which means that the replications are not necessarily estimating different true effect sizes \cite{button2013power}. 

Again in our first example, we did not find statistical heterogeneity at experiment level. However, failure to find statistical heterogeneity at experiment level with a specific estimation method (REML~\cite{langan2018comparison} in our case) does not imply that statistical heterogeneity did not materialize. In particular, if we had used another estimation method to estimate $\tau^2$ instead (e.g., the Der Simonian and Laird estimator \cite{langan2018comparison}), $\tau^2$ may not have been 0---and, thus, statistical heterogeneity may have materialized. Thus, a possible safeguard is to use different methods to estimate $\tau^2$ before making definite claims on whether or not statistical heterogeneity materialized across the replications \cite{langan2018comparison}.

In our second example, we did find statistical heterogeneity at experiment level, and subsequently at participant level. This shows that heterogeneity may also materialize if unacknowledged characteristics of the experimental configurations or the participants (as in this example) affect the results---despite the best efforts of the experimenters to keep the characteristics of the replications as identical as possible. Particularly, as the results of SE experiments depend upon a myriad of contextual variables \cite{basili1999building}, this possibility cannot be ruled out. This means that it is important to always check for heterogeneity, as even a priori identical replications may provide heterogeneous results.   

In our third example, we found statistical heterogeneity at experiment level which was then traced to configuration level. Note that it would be helpful to try to explore experimental configurations prospectively.

In all our examples, given the imprecision of results regarding 95\% CI for effect sizes and $I^2$, a key issue is to try to run a priori power analyses for prospective studies, and run more replications to incorporated them to the meta-analysis.

Finally, we suggest that the results of SE experiments be regarded not as the absolute truth that needs to be replicated and verified but, instead, as a partial view of the results within the population. This is due to the fact that small sample sizes are very common in SE\footnote{Note that this a sub-optimal approach because of the threat of introducing heterogeneity due to unacknowledged variables. It is better to conduct fewer larger studies. Very often, however, the only option is to run several small studies.} \cite{dybaa2006systematic}\cite{adrisms}. Only after collecting large chunks (i.e., running large-scale experiments) or assembling many\footnote{Unfortunately, there are no hard-and-fast rules for establishing how many replications are enough \cite{borenstein2011introduction}. This is because the precision of the results may be affected by the distribution of sample sizes across the replications \cite{ruvuna2004unequal}, the experimental design of the replications \cite{morris2002combining}, the variability of the data \cite{cumming2013understanding}, etc.} small pieces of evidence by applying the appropriate statistical methods (meta-analysis) is it possible to see the big picture in all its detail. From this perspective, even the tiniest of replications is key in SE---as long as the replication is of comparable quality to the baseline experiment, and the results of the baseline and the replication are meta-analyzed together.

As a summary, we provide a step-by-step guide to verify experimental results in SE. It is shown in Figure~\ref{guide}.


\begin{figure}[h!]  
  \caption{A step-by-step guide to verify experimental results in SE.}
  \label{guide}
  \centering
  \noindent\fbox{
   \parbox{4in}{
      \begin{enumerate}
    
         \item Use random-effects meta-analysis (instead of focusing on reproducibility of p-values or effect sizes).

         \item Interpret results in terms of calculated joint effect size and its associated 95\% CI (instead of regarding the results of each experiment individually).

         \item Check for possible sample variables influencing the results using linear mixed models (note that the raw data of all experiments is needed and the sample variables should have been measured).
         
         \item Check for heterogeneity by interpreting the value of $I^2$ and its 95\% CI.
         
         \item If heterogeneity has materialized, check the possible influence of configuration variables using sub-group meta-analysis.
         
         \item If the heterogeneity value is inaccurate (wide $I^2$ CI), take into consideration that more experiments are needed.
         
         \item Keep in ind that causality cannot be derived from the configuration and sample variables analyzed, even when a statistical relationship has been identified.

       \end{enumerate}
  }
}
\\

\end{figure}


\section{Threats to validity}
\label{threats}

We acknowledge that other researchers might have gathered a different list of references portraying the relevance and role of replication, and that this may have biased our findings. Unfortunately, we were unable to gather these references systematically because the search of online databases using the terms \say{replication} and \say{experiments} returned an  the unmanageable list of references. However, we tried to minimize the impact of this shortcoming by relying on well-known references in replication in other disciplines \cite{thompson1994pivotal}\cite{pashler2012editors}\cite{baker2016there}\cite{borenstein2011introduction}\cite{egger2008systematic}\cite{ioannidis2008research}\cite{cumming2014new} and systematic mapping studies on replication in SE \cite{de2015investigations}\cite{da2014replication}\cite{bezerra2015replication}. Additionally, we complemented our findings with simulations, and the results of three real-life groups of replications from our own research. We trust that this approach led to findings representative of the current state of the art on replication---at least in SE.

How representative are the results of normal data, parametric tests and parametric effect sizes for the simulations for SE? Normal data are not common in SE \cite{arcuri2011practical}\cite{kitchenham2017robust}. Thus, other data distributions (e.g., beta distributions with different degrees of skewness, non-central t-distributions, etc. \cite{cumming2013understanding}) might have been more representative for our simulations. However, we resorted to normal distributions to conduct the simulations because SE data follow myriad distributions \cite{kitchenham2017robust}\cite{arcuri2011practical},  Cohen's d is commonly used to synthesize the results of SE experiments \cite{kampenes2007systematic} and  Cohen's d is simple to calculate with normal distributions \cite{cumming2013understanding}. We acknowledge that if we had simulated replications with other data distributions (e.g., skewed beta distributions), we may have had to rely on other effect sizes (e.g., Cliff's delta etc. \cite{kitchenham2017robust}) to analyze the data. In turn, this may have affected the probability of finding similar results in terms of $p$-values and effect sizes. Regardless of the data distributions that we simulated, our simulations were not in any case designed to show which statistical tests---or effect sizes---are potentially the best for analyzing the data---as this topic has already been addressed elsewhere \cite{kitchenham2017robust}\cite{arcuri2011practical}. Instead, the aim of our simulations was to illustrate how experiments are just samples from a larger population, and, as such, their results are largely dependent upon their sample sizes---and the true effect size being estimated. If we had simulated replications with other data distributions, their results would have also fluctuated around the true effect size in the population \cite{cumming2013understanding}. From this point of view, we do not expect our findings to be limited by the data distributions that we selected.

\section{Conclusions and Future Work}
\label{conclusion}

It has been argued that the purpose of SE replications is to verify the results of previous experiments \cite{de2015investigations}\cite{da2014replication}\cite{bezerra2015replication}. Different methods have been proposed to achieve this aim. Some of techniques overlook the problem of sampling error, aggravated by small sample sizes which are common in SE experiments. In turn, even replications estimating an identical true effect size may provide conflicting results (e.g., negative and positive results), although this does not mean that results are really contradictory \cite{button2013power}.

In view of this, we argue that p-values and effect sizes are unsuitable for assessing whether or not the results of previous experiments hold. This is because sampling error may be masking the true effect size being estimated in each replication. SE replications need to go hand-by-hand with meta-analysis.
 
Additionally, contextual variables may be impacting  the results of SE experiments \cite{basili1999building}\cite{wohlin2012experimentation}. The purposeful or inadvertent manipulation of the experimental configurations or sample characteristics of the replications may or may not alter the true effect size being estimated in each replication. Meta-analysis can help to assess whether there are contextual variables affecting the results.

Meta-analysis is the glue that keeps the results of the replications together. From this point of view, the results achieved in small baseline experiments or replications are no longer regarded as the absolute truth that needs to be replicated, but, instead, as a small piece of evidence within a larger picture that only emerges after assembling many small pieces  \cite{cumming2014new}. 

These results are useful for empirical software engineers and software developers. They suggest that empirical software engineers should be aware of the limitations of using p-values and effect sizes for checking the results of replications. They should also know the potential of meta-analysis. Lastly, they should acknowledge the importance of context when running replications. On the other hand, our results suggest that software developers should take into consideration the method used to check the results of previous experiments when using the evidence generated by empirical software engineers. Additionally, they should learn to understand the impact of context on the evidence generated by empirical software engineers.

Replication is not yet well understood by the SE community. There are still many issues that need to be investigated, including, for example, whether different replication types play different roles; how to estimate the number of replications needed to output a reliable piece of knowledge; which changes should be made to a baseline experiment when running replications, as well as the order in which these changes should be made; which are the best protocols for running replications (e.g. whether multiple replications should be run by different researchers at the same time or successively). First and foremost, however, we must learn how to encourage researchers to run replications in SE and raise awareness of how just important they are.

\begin{acknowledgements} 
This research was developed with the support of project PGC2018-097265-B-I00, funded by: FEDER/Spanish Ministry of Science and Innovation---Research State Agency.

\end{acknowledgements}

\bibliographystyle{spmpsci}      
\bibliography{biblio}   

%
%

\begin{appendices}

\section{Probability of Reproducing the Results of Previous Experiments}
\label{reproducing_results}

\begin{figure}[h!]  
  \caption{Probability that all replications reproduce the results of previous experiments by means of $p$-values in groups of 2 to 12 replications, all estimating a true Cohen's of 0.5.}
  \label{identical_results_medium}
  \centering
    \includegraphics[width=\textwidth,keepaspectratio]{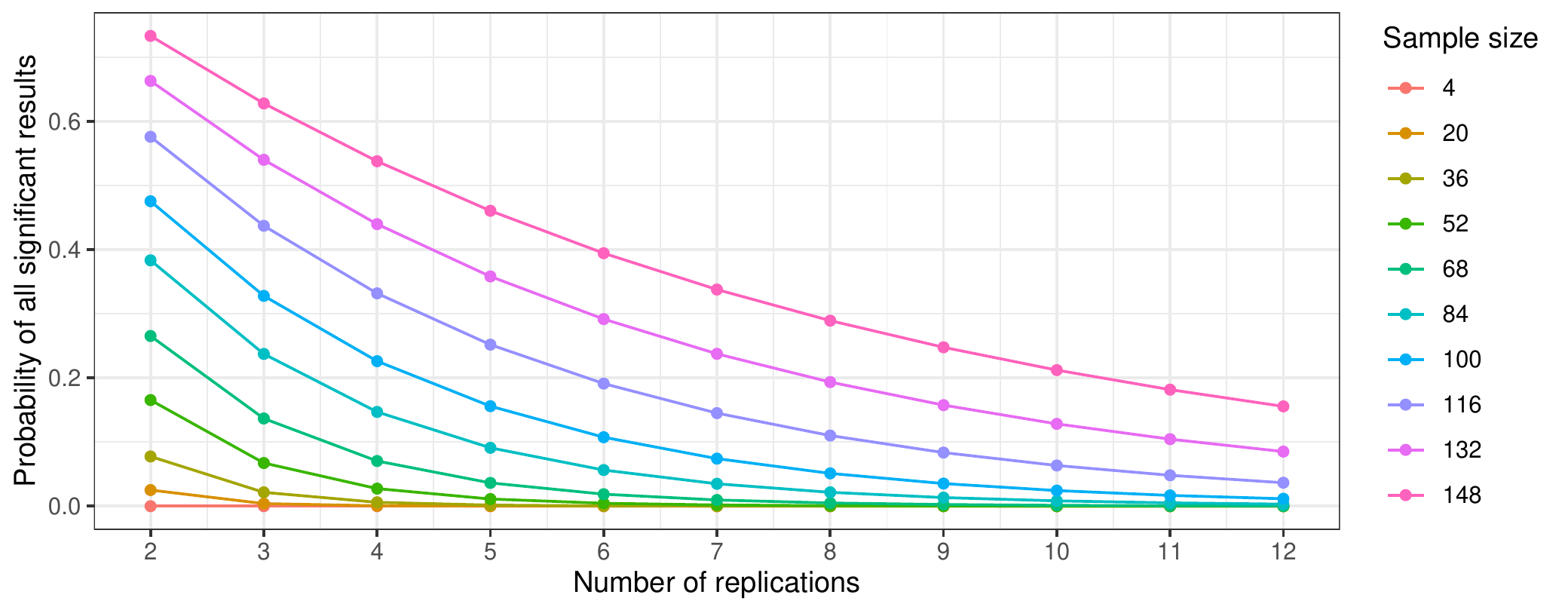}
\end{figure}

Figure~\ref{identical_results_medium} shows the probability of reproducing the results of previous experiments by means of $p$-values in groups of 2 to 12 replications for a true Cohen's d of 0.5 at different sample sizes. 

\begin{figure}[h!]  
  \caption{Probability that all replications reproduce the results of previous experiments by means of $p$-values in groups of 2 to 12 replications, all estimating a true Cohen's of 0.8.}
  \label{identical_results_large}
  \centering
    \includegraphics[width=\textwidth,keepaspectratio]{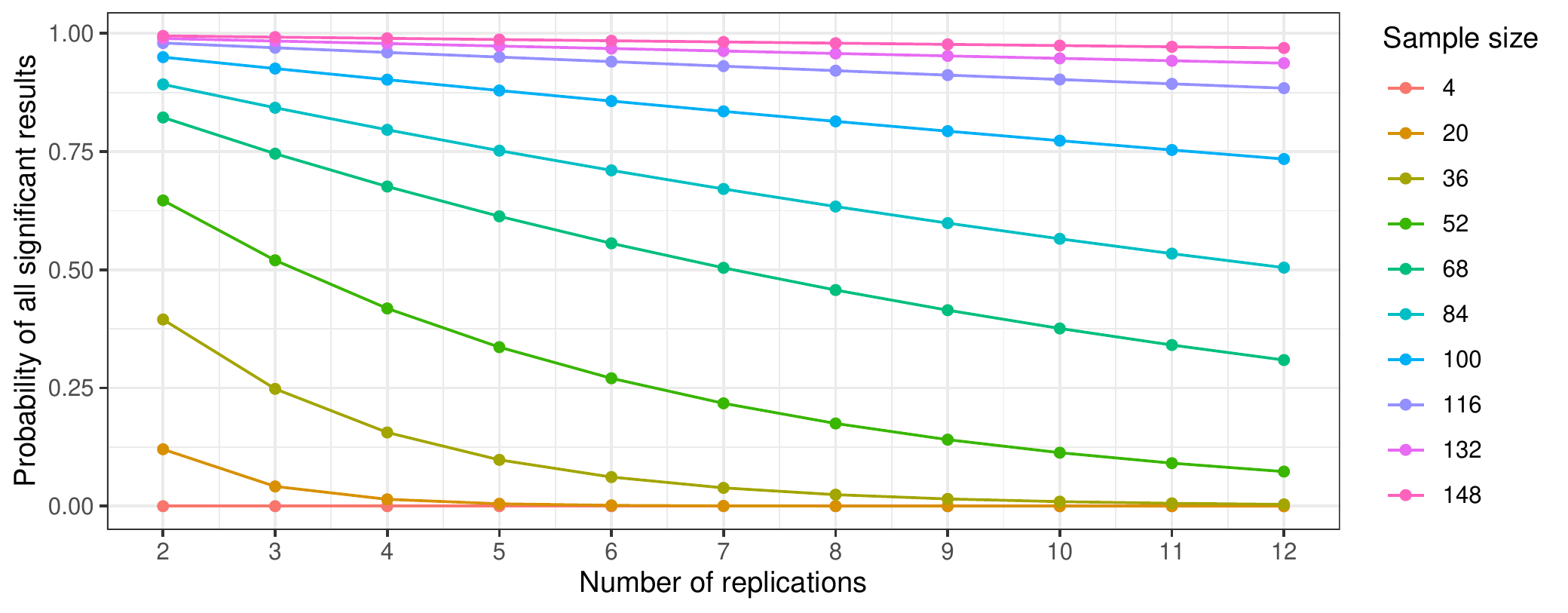}
\end{figure}

Figure~\ref{identical_results_large} shows  the probability of reproducing the results of previous experiments by means of $p$-values in groups of 2 to 12 replications for a true Cohen's d of 0.8 at different sample sizes.

\section{Joint Results of MA vs. Overall Result of the Large-Scale Experiment}
\label{chunks}

\begin{figure}[h!]  
  \caption{Joint result of meta-analysis vs. overall result of the large-scale experiment.}
  \label{chunks_big}
  \centering
  \includegraphics[width=\textwidth,keepaspectratio]{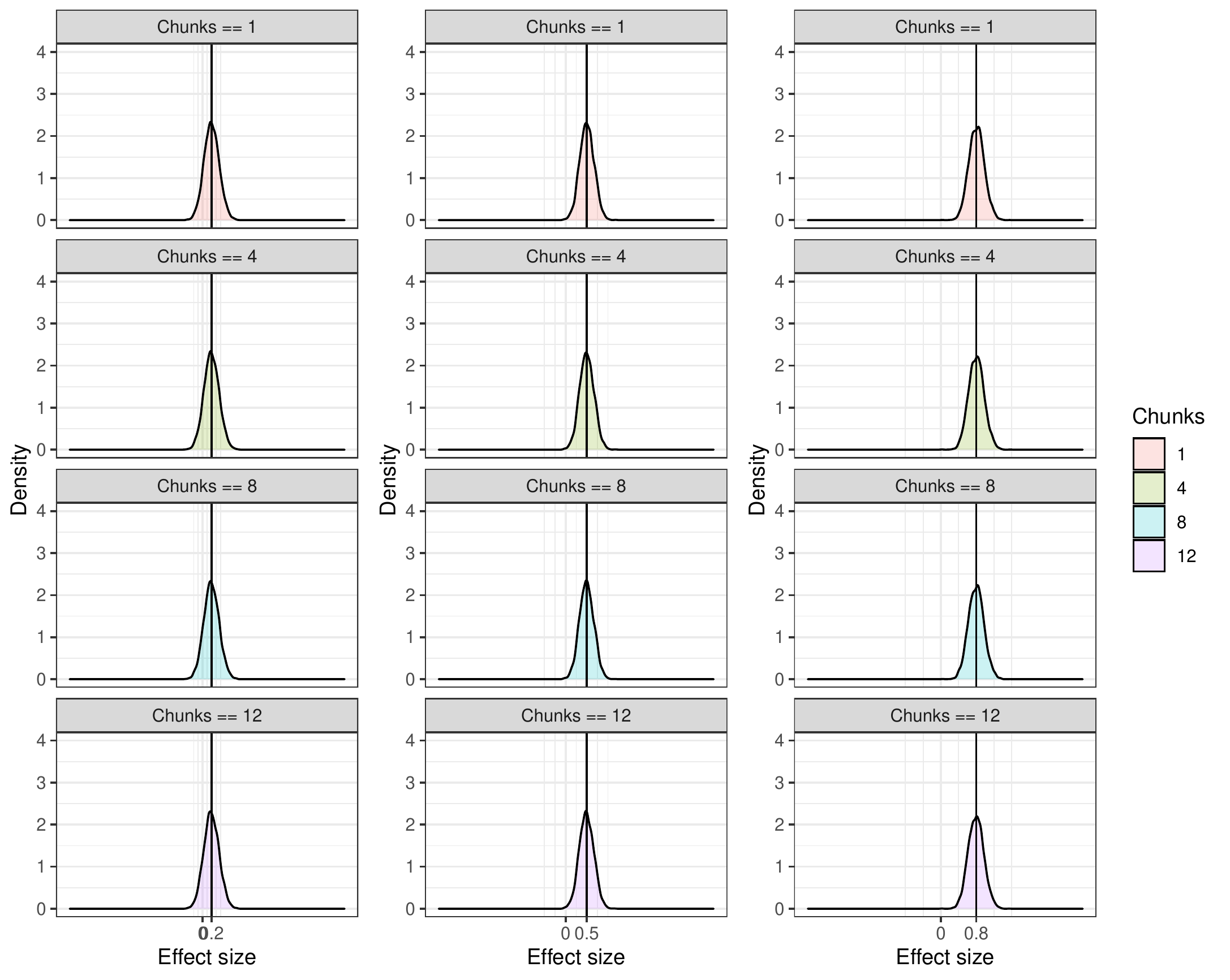}
\end{figure}

Figure~\ref{chunks_big} shows the joint estimated effect sizes of meta-analysis for 4, 8 and 12 chunks of sample sizes 36, 18 and 12 each, versus the estimated effect size of the large-scale experiment of sample size 144 at true Cohen's d of 0.2, 0.5 and 0.8.

\section{Effect Size Variability when Meta-Analyzing Experiments}
\label{variability_effect_sizes}

\begin{figure}[h!]  
  \caption{Meta-analysis: groups with different numbers of replications and true effect sizes and a sample size of 4.}
  \label{meta_analysis_4}
  \centering
    \includegraphics[width=\textwidth,keepaspectratio]{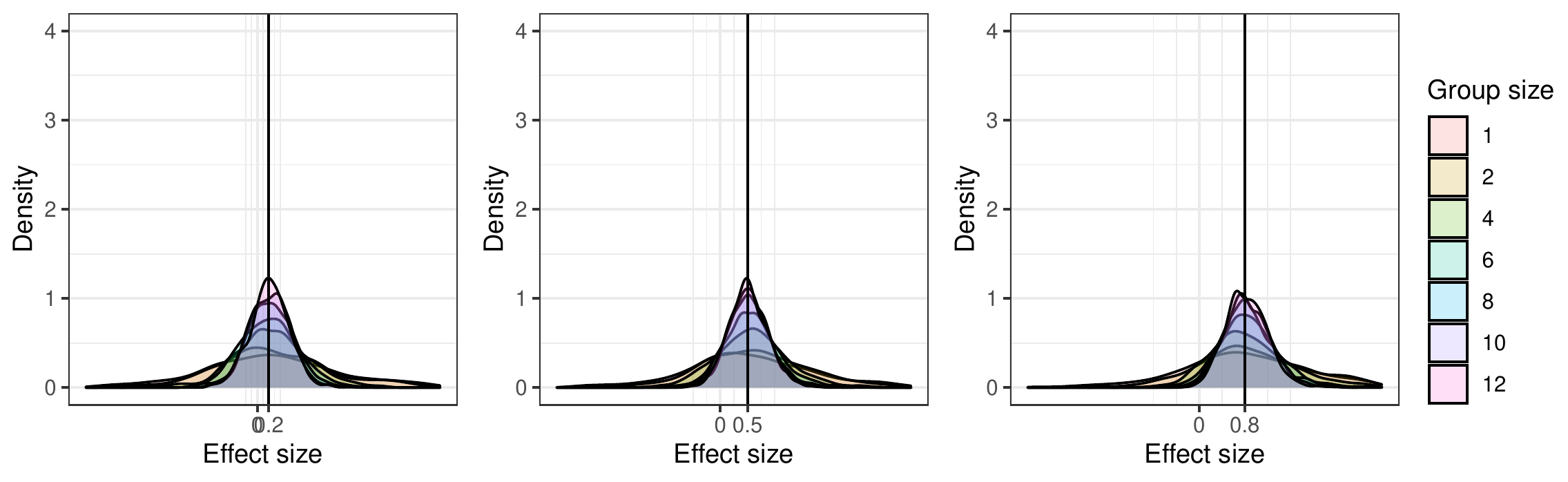}
\end{figure}

Figure~\ref{meta_analysis_4} shows the distribution of joint effect sizes achieved in 5,000 groups of 1 to 12 identical AB between-subjects replications each with a sample size of 4, meta-analyzed at different true effect sizes (i.e., Cohen's d of 0.2, 0.5 and 0.8, corresponding to a small, medium and large true effect size, from left to right, respectively). 

Table~\ref{confidence_intervals_3} shows the 95\%CIs for the joint effect sizes estimated in Figure~\ref{meta_analysis_4}.

\begin{table}[h!] \centering 
  \caption{95\% confidence intervals for joint estimated effect sizes} 
  \label{confidence_intervals_3} 
\begin{tabular}{@{\extracolsep{5pt}} ccccc} 
\\[-1.8ex]\hline 
\hline \\[-1.8ex] 
True & Group & \multicolumn{3}{c}{Joint estimated effect size} \\ \cline{3-5} 
effect size & size & Lower bound 2.5\% & Mean & Upper bound 97.5\% \\ 
\hline \\[-1.8ex] 
$0.2$ & $1$ & $$-$3.79$ & $0.40$ & $5.65$ \\ 
 & $2$ & $$-$2.18$ & $0.36$ & $3.80$ \\ 
 & $4$ & $$-$1.14$ & $0.26$ & $1.63$ \\ 
 & $6$ & $$-$0.72$ & $0.22$ & $1.25$ \\ 
 & $8$ & $$-$0.63$ & $0.23$ & $1.08$ \\ 
 & $10$ & $$-$0.47$ & $0.22$ & $0.93$ \\ 
 & $12$ & $$-$0.48$ & $0.22$ & $0.90$ \\ 
$0.5$ & $1$ & $$-$3.09$ & $0.76$ & $5.89$ \\ 
 & $2$ & $$-$1.54$ & $0.83$ & $4.19$ \\ 
 & $4$ & $$-$0.65$ & $0.59$ & $1.98$ \\ 
 & $6$ & $$-$0.37$ & $0.57$ & $1.72$ \\ 
 & $8$ & $$-$0.27$ & $0.55$ & $1.47$ \\ 
 & $10$ & $$-$0.17$ & $0.51$ & $1.29$ \\ 
 & $12$ & $$-$0.16$ & $0.51$ & $1.23$ \\ 
$0.8$ & $1$ & $$-$2.05$ & $1.42$ & $6.71$ \\ 
 & $2$ & $$-$0.99$ & $1.24$ & $5.02$ \\ 
 & $4$ & $$-$0.24$ & $0.94$ & $2.81$ \\ 
 & $6$ & $$-$0.11$ & $0.89$ & $2.17$ \\ 
 & $8$ & $0.07$ & $0.85$ & $1.90$ \\ 
 & $10$ & $0.12$ & $0.85$ & $1.70$ \\ 
 & $12$ & $0.18$ & $0.83$ & $1.60$ \\ 
\hline \\[-1.8ex] 
\end{tabular} 
\end{table} 

\begin{figure}[h!]  
  \caption{Meta-analysis: groups with different numbers of replications and true effect sizes and a sample size of 100.}
  \label{meta_analysis_100}
  \centering
    \includegraphics[width=\textwidth,keepaspectratio]{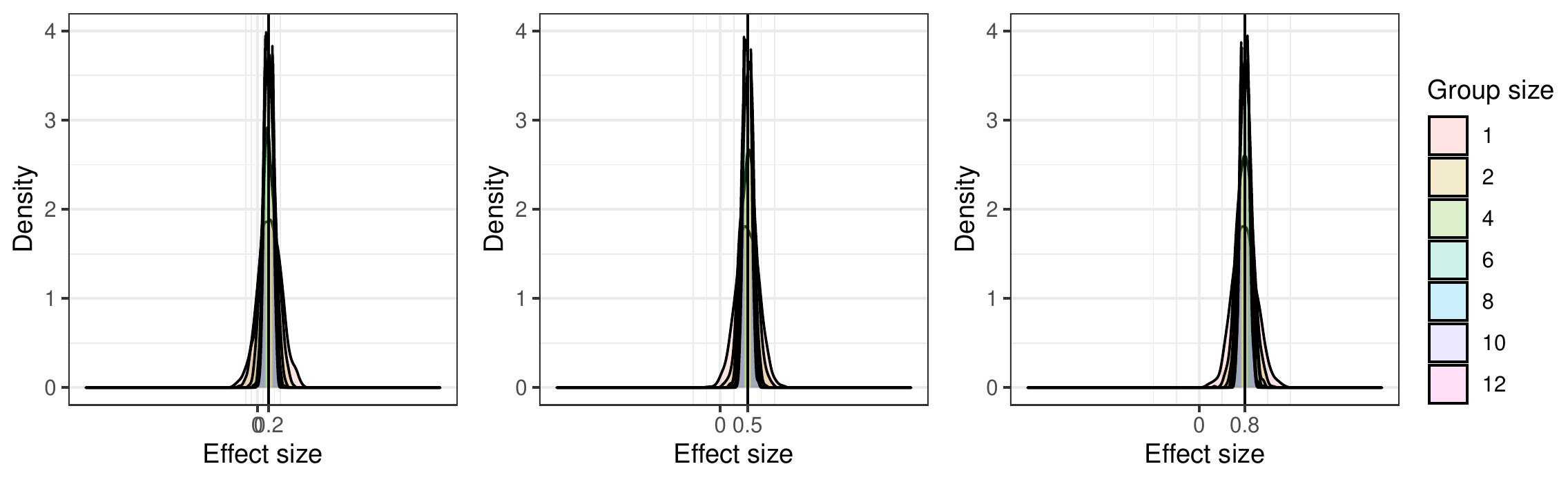}
\end{figure}

Figure~\ref{meta_analysis_100} shows the distribution of joint effect sizes achieved in 5,000 groups of 1 to 12 identical AB between-subjects replications each with a sample size of 100, meta-analyzed at different true effect sizes (i.e., Cohen's d of 0.2, 0.5 and 0.8, corresponding to a small, medium and large true effect size, from left to right, respectively). 

Table~\ref{confidence_intervals_4} shows the 95\%CIs for the joint effect sizes estimated in Figure~\ref{meta_analysis_100}.

\begin{table}[h!] \centering 
  \caption{95\% confidence intervals for joint estimated effect sizes} 
  \label{confidence_intervals_4} 
\begin{tabular}{@{\extracolsep{5pt}} ccccc} 
\\[-1.8ex]\hline 
\hline \\[-1.8ex] 
True & Group & \multicolumn{3}{c}{Joint estimated effect size} \\ \cline{3-5} 
effect size & size & Lower bound 2.5\% & Mean & Upper bound 97.5\% \\ 
\hline \\[-1.8ex] 
$0.2$ & $1$ & $$-$0.19$ & $0.22$ & $0.63$ \\ 
 & $2$ & $$-$0.12$ & $0.20$ & $0.49$ \\ 
 & $4$ & $$-$0.02$ & $0.20$ & $0.41$ \\ 
 & $6$ & $0.06$ & $0.21$ & $0.36$ \\ 
 & $8$ & $0.07$ & $0.20$ & $0.34$ \\ 
 & $10$ & $0.08$ & $0.20$ & $0.33$ \\ 
 & $12$ & $0.09$ & $0.20$ & $0.32$ \\ 
$0.5$ & $1$ & $0.09$ & $0.50$ & $0.89$ \\ 
 & $2$ & $0.24$ & $0.51$ & $0.80$ \\ 
 & $4$ & $0.31$ & $0.50$ & $0.70$ \\ 
 & $6$ & $0.34$ & $0.50$ & $0.67$ \\ 
 & $8$ & $0.34$ & $0.50$ & $0.65$ \\ 
 & $10$ & $0.37$ & $0.50$ & $0.64$ \\ 
 & $12$ & $0.39$ & $0.50$ & $0.62$ \\ 
$0.8$ & $1$ & $0.41$ & $0.80$ & $1.23$ \\ 
 & $2$ & $0.53$ & $0.80$ & $1.10$ \\ 
 & $4$ & $0.60$ & $0.81$ & $1.01$ \\ 
 & $6$ & $0.63$ & $0.80$ & $0.99$ \\ 
 & $8$ & $0.66$ & $0.80$ & $0.95$ \\ 
 & $10$ & $0.68$ & $0.80$ & $0.93$ \\ 
 & $12$ & $0.68$ & $0.80$ & $0.92$ \\ 
\hline \\[-1.8ex] 
\end{tabular} 
\end{table}

\end{appendices}

\clearpage

\section*{Biographies}


\hfill \break
\begin{floatingfigure}{1.2in}
\includegraphics[width=0.20\textwidth,clip,keepaspectratio]{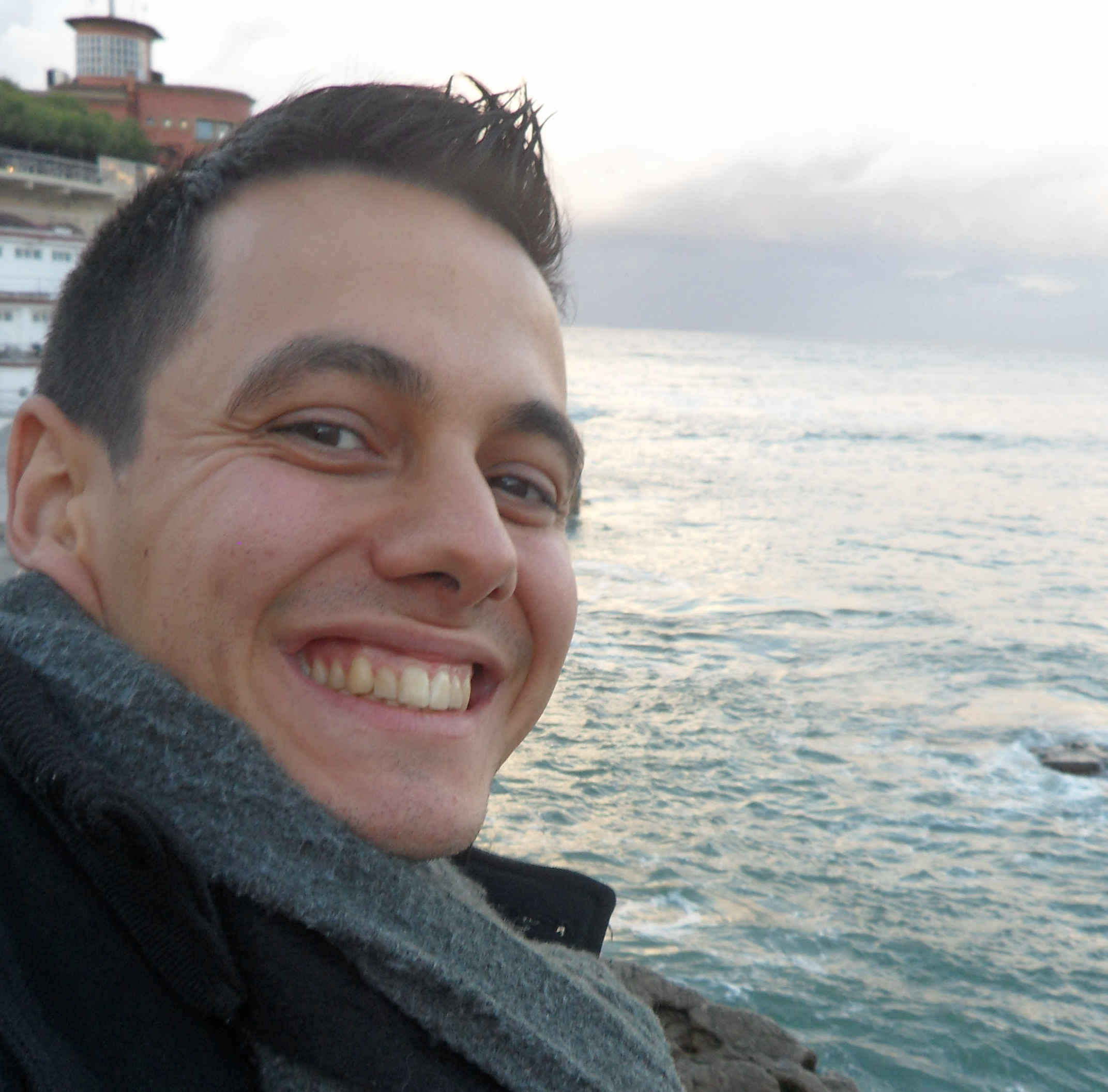}
\end{floatingfigure}
\noindent \textbf{Adrian Santos} received his MSc in Software and Systems and MSc in Software Project Management at the Technical University of Madrid, Spain, and his MSc in IT Auditing, Security and Government at the Autonomous University of Madrid, Spain. He obtained his PhD in Software Engineering at the University of Oulu, Finland. He is currently working as a software engineer in industry. 


\hfill \break
\begin{floatingfigure}{1.2in}
\includegraphics[width=0.20\textwidth,clip,keepaspectratio]{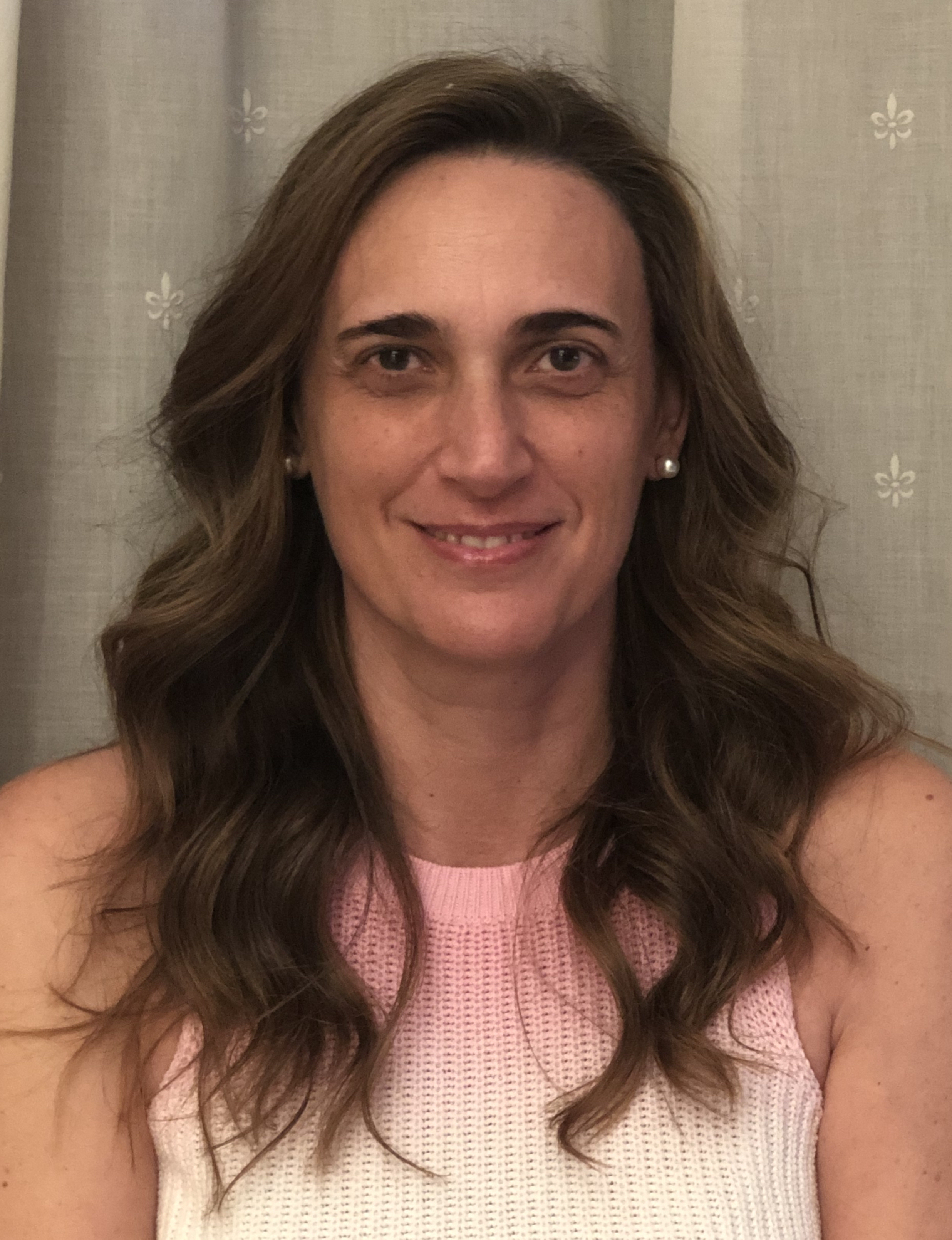} 
\end{floatingfigure}
\noindent \textbf{Sira Vegas} has been associate professor of software engineering with the  School of Computer Engineering at the Technical University of Madrid, Spain, since 2008. Sira belongs to the review board of IEEE Transactions on Software Engineering, and is a regular reviewer of the Empirical Software Engineering Journal. She was program chair for the International Symposium on Empirical Software Engineering and Measurement in 2007.


\hfill \break
\begin{floatingfigure}{1.2in}
\includegraphics[width=0.20\textwidth,clip,keepaspectratio]{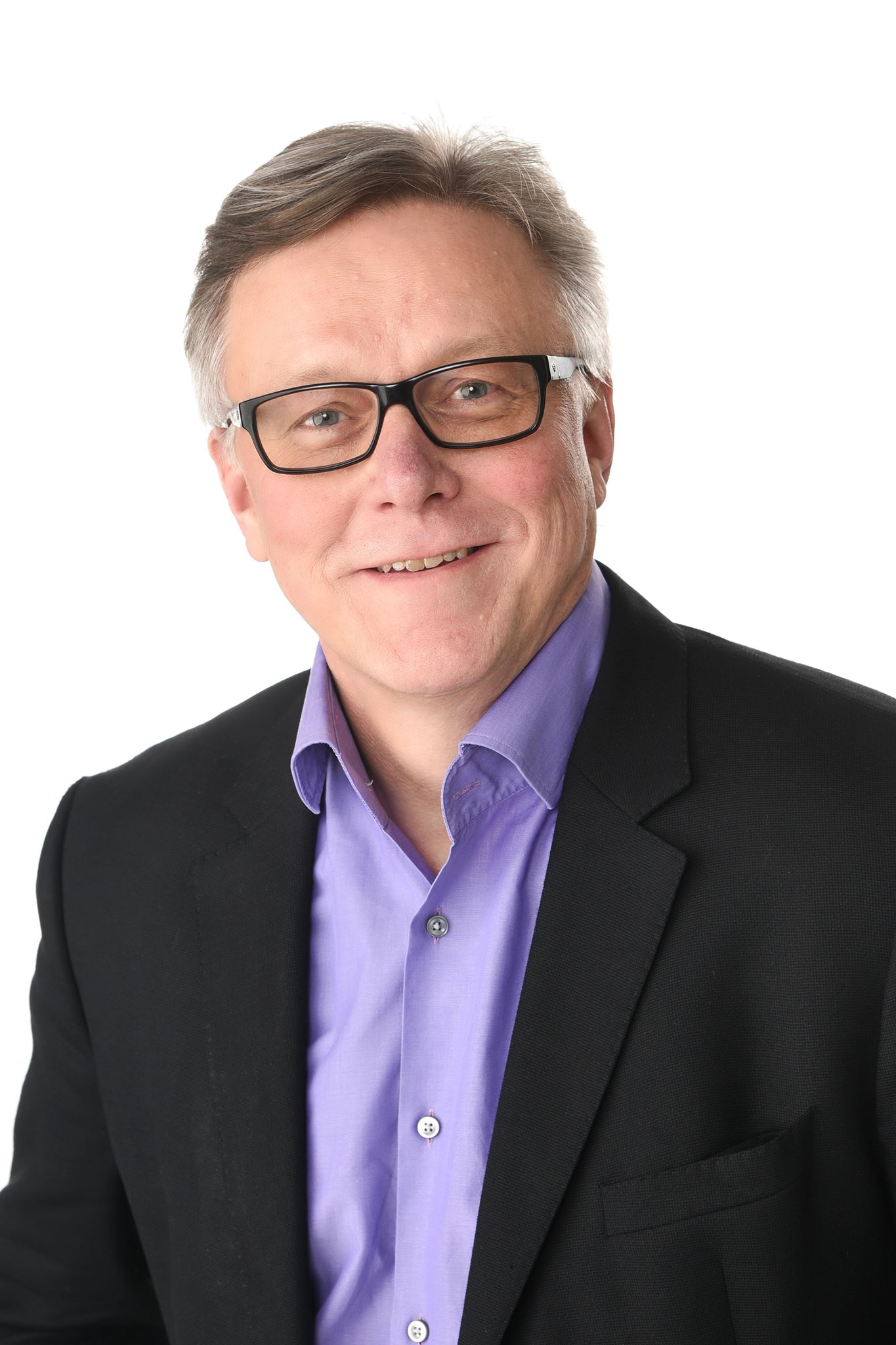} 
\end{floatingfigure}
\noindent \textbf{Markku Oivo} is professor and head of the M3S research unit at the University of Oulu, Finland. He had visiting positions at the University of Maryland (1990-91), Schlumberger Ltd. (Paris 1994-95), Fraunhofer IESE (1999-2000), University of Bolzano (2014-15), and Universidad Polit\'ecnica de Madrid (2015). 
\newline

\vspace{1cm}

\hfill \break
\begin{floatingfigure}{1.2in}
\includegraphics[width=0.20\textwidth,clip,keepaspectratio]{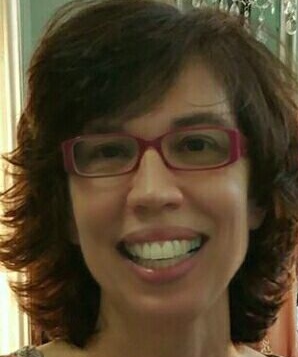} 
\end{floatingfigure}
\noindent \textbf{Natalia Juristo}  has been full professor of software engineering with the  School of Computer Engineering at the Technical University of Madrid, Spain, since 1997. She was awarded a FiDiPro (Finland Distinguished Professor Program) professorship at the University of Oulu, from January 2013 until June 2018. Natalia belongs to the editorial board of EMSE and STVR. In 2009, Natalia was awarded an honorary doctorate by Blekinge Institute of Technology in Sweden.



\end{document}